\documentclass{article}

\usepackage{amsmath}
\usepackage{amssymb}
\usepackage{graphicx}
\usepackage{floatrow}
\usepackage{dcolumn}
\usepackage{bm}
\usepackage{booktabs}
\usepackage{adjustbox}
\usepackage{authblk}

\usepackage[utf8]{inputenc}
\usepackage[T1]{fontenc}
\usepackage{mathptmx}
\usepackage{etoolbox}
\usepackage{subcaption}
\usepackage{units}

\newcommand\ddfrac[2]{\frac{\displaystyle #1}{\displaystyle #2}}
\newcommand{\xn}{\hat{x}}
\newcommand{\yn}{\hat{y}}
\newcommand{\zn}{\hat{z}}
\newcommand{\cn}{\hat{c}}
\newcommand{\tn}{\hat{t}}
\newcommand{\vn}{\hat{v}}
\newcommand{\pn}{\hat{p}}
\newcommand{\an}{\hat{a}}

\makeatletter
\def\@email#1#2{%
 \endgroup
 \patchcmd{\titleblock@produce}
  {\frontmatter@RRAPformat}
  {\frontmatter@RRAPformat{\produce@RRAP{*#1\href{mailto:#2}{#2}}}\frontmatter@RRAPformat}
  {}{}
}
\makeatother
\begin{document}

\title{Solution approaches for evaporation-driven density instabilities in a slab of saturated
porous media}

\author[1]{Leon H. Kloker}
\author[1]{Carina Bringedal}

\affil[1]{Institute for Modelling Hydraulic and Environmental Systems, University of Stuttgart}
\affil[ ]{carina.bringedal@iws.uni-stuttgart.de}

\date{}

\maketitle

\begin{abstract}
This work considers the gravitational instability of a saline boundary layer formed by an evaporation-induced flow through a fully-saturated porous slab. Soil salinization appears in an increasing amount of regions and often adversely impacts biological activity. In extreme cases, evaporation of saline waters can eventually result in the formation of salt lakes as salt accumulates. Modeling such a natural system is of paramount importance in order to understand and obstruct the processes leading to the build-up of salt. As natural convection can impede the accumulation of salt, establishing a relation between its occurrence and the value of physical parameters such as evaporation rate, height of the slab or porosity is crucial. One step towards determining when gravitational instabilities can arise is to compute the ground-state salinity, that evolves due to the uniform upwards flow caused by evaporation. The resulting salt concentration profile exhibits a sharply increasing salt concentration in the vicinity of the surface, which can lead to a gravitationally unstable setting. In this work, this ground state is analytically derived within the framework of Sturm-Liouville theory. Then, the method of linear stability in conjunction with the quasi-steady state approach (QSSA) - also called frozen profile approach - is employed to investigate the occurrence of instabilities. These instabilities can develop and grow over time depending on the Rayleigh number and the dimensionless height of the porous medium. In order to calculate the critical Rayleigh number, which can be used to determine the stability of a particular system, the eigenvalues of the linear perturbation equations have to be computed. Here, a novel fundamental matrix method is proposed to solve this perturbation eigenvalue problem and shown to coincide with an established Chebyshev-Galerkin method in their shared range of applicability. Finally, a 2-dimensional direct numerical simulation of the full equation system via the finite volume method is employed to validate the time of onset of convective instabilities predicted by the linear theory. Moreover, the fully nonlinear convection patterns, in this context usually referred to as salt fingers, are analyzed.
\end{abstract}

\section{\label{introduction}Introduction}
Evaporation of saline water from soils can be a cause for soil salinization which impedes plant growth and, thus, makes vast areas unreclaimable for agricultural purposes \cite{daliakopoulos2016threat}. During the evaporation process, salt is left behind, resulting in a gradual build-up of the salt concentration near the top surface of the soil. Once the salt concentration exceeds its solubility limit, salt precipitates. The resulting salt crust has an even more extreme impact on biological activity and can eventually result in erosion of the soil \cite{singh2016microbial, mejri2020modeling}. As an increasing number of regions are adversely affected by this problem \cite{okur2020soil, vereecken2009research}, the study of mechanism that are able to prevent salinization is gaining interest.

While water leaves the soil at the surface due to evaporation, salt dissolved in the water stays behind and, hence, accumulates over time \cite{allison1985estimation}. This, however, leads to an increasing density of the remaining liquid since the liquid density rises with its salinity \cite{geng2015numerical}. Because the salt has mainly accumulated near the top of the soil, where the water evaporates, the water near the surface gets gradually denser. The resulting density stratified setting may be unstable as the dense layer of liquid at the surface is subject to greater gravitational forces than the underlying layers \cite{wooding1997convection, gilman1996influence}. Under the appropriate conditions, instabilities are growing in the form of salt fingers \cite{chen1988onset, wooding1997convection2}, which start at the surface and expand downwards. Such fingers transport the accumulated salt from the upper to the lower part of the soil, where the salt concentration is smaller. The occurrence of salt fingers can be quantified by a dimensionless Rayleigh number, where numbers above a threshold - the critical Rayleigh number \cite{sutton1970onset} - mean that an infinitesimal perturbation will grow and induce convection. Hence, this is a natural mechanism which can hinder the precipitation of salt at the surface due to the development of convective instabilities, as the salt fingers can prevent the salt concentration at the surface from increasing any further. Consequently, determining under which conditions convective instabilities can occur is paramount to the understanding of soil salinization and its prevention \cite{shokri2020water}.

The stability of a particular solution of a system of differential equations can be investigated by a linear stability analysis \cite{nield2017internal}.
Such an analysis has already been successfully applied to porous media problems in several studies, e.g.~by Wooding et al. \cite{wooding1997convection}, van Duijn et al. \cite{van2001stability} and Pieters et al. \cite{pieters2018stability}, in order to quantify the occurrence of convective instabilities in groundwater flow caused by varying salt concentrations. These studies, however, consider a prescribed density or salinity at the top of the domain. Hence, it is either assumed that the system is coupled to a reservoir of fixed fluid density or the solubility limit of salt has already been reached at the surface. However, during soil salinization or the formation of salt lakes, the salt concentration increases over time as it is left behind during evaporation. To investigate convective instabilities before salt starts to precipitate, Gilman et al. \cite{gilman1996influence} introduced a Robin type, no-flux condition for salt at the top boundary. They employ this boundary condition in order to model the salt accumulation during evaporation from the vadoze zone in a vertically bounded domain. Bringedal et al. \cite{bringedal2022evaporation} have successfully employed the same boundary condition to model evaporation of saline water from a fully saturated, vertically unbounded porous medium.

In this paper, we consider a vertically bounded porous slab, and assume, as was also done in the work of Bringedal et al. \cite{bringedal2022evaporation}, that the entire porous slab is fully saturated with water. This holds true as long as the capillary pressure remains below the entry pressure of the soil. Even though this is generally not the case since evaporation commonly results in an unsaturated zone in the vicinity of the surface, considering single-phase flow will allow a simplified analysis whose results still give a better understanding of the development of instabilities in the described scenario.

Nevertheless, the problem which we consider here differs in several ways from the studies by Wooding et al. \cite{wooding1997convection}, van Duijn et al. \cite{van2001stability} and Pieters et al. \cite{pieters2018stability} as a Robin boundary condition is used to model the salt behaviour before the solubility limit is reached. In contrast to the work done by Bringedal et al. \cite{bringedal2022evaporation}, a porous layer with finite height is considered, which, especially for small heights, strongly changes the ground-state salt concentration profile and thus also the stability bound. 

In this setting, we expect salt to accumulate at the top of the porous medium over time while diffusion leads to a more even distribution within the domain. As the salt concentration and therefore the density difference between the bottom and top of the porous medium increase with time, the flow is more prone to develop convective instabilities. The linear stability analysis will yield an estimate of the critical Rayleigh number at a given time and for a certain height of the domain by solving the corresponding perturbation eigenvalue problem.

The resulting eigenvalue problem can generally not be solved explicitly, but relies on numerical approximations. Previous studies have employed several different numerical methods to solve the eigenvalue problem arising from the perturbation equations. For example, Gilman et al. \cite{gilman1996influence} used a finite difference method, van Duijn et al. \cite{van2001stability} utilized the Jacobi-Davidson scheme, Chan et al. \cite{chan2012linear} employed a spectral method and Bringedal et al. \cite{bringedal2022evaporation} as well as Pieters et al. \cite{pieters2018stability} applied a Galerkin or Petrov-Galerkin approach to solve the eigenvalue problem. In this study, however, we introduce a new approach to solve the eigenvalue problem, which can be applied not only to the specific considered setting, but to general perturbation eigenvalue problems. Here, the idea is to use a property of the fundamental matrix of the perturbation equations in order to determine if non-trivial solutions exist. Additionally, a Chebyshev-Galerkin method is used to create reference solutions to verify the results obtained by the novel method as well as to complement its range of applicability.

The paper is structured as follows: In section \ref{mathematical_model}, the mathematical model of our considered setup is introduced. This comprises the model equations that are stated in subsection \ref{Domain_model_equations} as well as initial and boundary conditions presented in subsection \ref{initial_boundary_conditions}. Section \ref{Stability_analysis} concerns the linear stability analysis. Here, the model is nondimensionalized in subsection \ref{nondimensional_model}, the ground state is calculated in subsection \ref{ground_state_solution}, the linear perturbation equations are derived in subsection \ref{linear_perturbation_equations} and the resulting eigenvalue problem is formulated in subsection \ref{eigenvalue_problem}. Finally, the new solution method as well as the already established Chebyshev-Galerkin scheme are introduced in subsection \ref{solution_methods} and the results of both methods are presented in subsection \ref{stability_limit}. In section \ref{numerical_solution}, the original mathematical model from section \ref{mathematical_model} is solved by a direct numerical simulation. This is done in order to validate the stability bounds predicted by the linear stability theory in subsection \ref{comparison_time_of_onset}, and to give insights into the nonlinear convection patterns in subsection \ref{full_convection_patterns} as the linear theory breaks down for larger instabilities. The final section \ref{conclusions} concludes the results of this study.

\section{\label{mathematical_model}Mathematical Model}

\subsection{\label{Domain_model_equations}Domain and Model equations}

We consider an isotropic, uniform slab of a porous medium which is mapped to the semi-infinite domain $\Omega_H = \{(x,y,z) \in \mathbb{R}^3 : 0 \leq z \leq H \}$, where $H$ is the height of the medium. The assumption that the entire domain is saturated allows us to model the problem with single-phase flow equations which enables the analytical treatment in the linear stability section \ref{Stability_analysis} later. The following equations govern the behaviour of the described system when the Boussinesq approximation is employed:
\begin{gather}
    \label{flow_conservation_special}
    \nabla \cdot \mathbf{v} = 0~, \\[2pt]
    \label{darcy_law}
    \mathbf{v} = -\dfrac{K}{\mu} \bigl(\nabla p + \rho g \mathbf{e_z} \bigr)~, \\[2pt]
    \label{salt_conservation}
    \phi \partial_t c + \nabla \cdot \bigl(c\mathbf{v}\bigr) = D \nabla^2 c~.
\end{gather}
Here, $\mathbf{v}$ is the Darcy flux, $K$ is the scalar permeability, $\mu$ is the dynamic viscosity of water and $p$ is the pressure. Moreover, the fluid density $\rho$, gravitational acceleration $g$, porosity $\phi$, effective diffusion constant $D$ of salt in water, as well as the salt concentration $c$ appear in the equations.

The first equation \eqref{flow_conservation_special} is the continuity equation under the Boussinesq approximation, which states that density differences of the fluid are small such that they can be neglected everywhere except when multiplied by the gravitational acceleration.
The second equation \eqref{darcy_law} is Darcy's law, the momentum equation describing fluid flow in porous media. Lastly, equation \eqref{salt_conservation} is a convection-diffusion equation governing the salt concentration. 

In this setting, we will neglect chemical reactions such as the precipitation of salt once the salinity exceeds the solubility limit. This means that the model together with the following investigations and results are only valid up to the solubility limit which is 359 gram of salt per liter \cite{burgess1978metal} at 20 degrees Celsius and atmospheric pressure. We also disregard the effect of temperature variations since they play only a minor role for the stability of the boundary layer in comparison to salinity differences in the considered scenario \cite{gilman1996influence}. Consequently, the third equation is coupled with Darcy's law via the linear equation of state
\begin{equation}
    \rho(c) = \rho_0 \bigl( 1 + \gamma (c-c_0) \bigr)~,
\end{equation}
which relates the water density only to the salinity $c$. In this equation, $\rho_0$ and $c_0$ are a reference density and salt concentration, and will be chosen as the initial density of the liquid and the initial salt concentration, respectively, which are both assumed constant in space. Also, $\gamma$ is the density expansion coefficient of water with increasing salinity.

\subsection{\label{initial_boundary_conditions}Initial and boundary conditions}

In the investigated scenario, the laterally unbounded porous slab is coupled to an infinite reservoir of saline groundwater with concentration $c_0$ at the bottom. At the surface, water is evaporating, e.g. due to wind or the influence of the sun. This is modeled by a fixed evaporation rate $E$ as vertical velocity. In order to adhere to the continuity equation, reservoir water has to enter the porous medium from below at the same rate. Since salt does not leave the porous slab during the evaporation process, a no-flux condition is imposed on the salinity at the top of the medium. In mathematical form, the boundary conditions of the system under investigation are given by:
\begin{align}
    \mathbf{v}(x,y,z=0,t) = \mathbf{v}(x,y,z=H,t) = E \mathbf{e_z}~, \\[3pt]
    c(x,y,z=0,t) = c_0 , ~\bigl( c\mathbf{v} - D \nabla c \bigr) \cdot \mathbf{e_z} \Big|_{z=H} = 0~.
    \label{salt_boundary_conditions}
\end{align}
As an initial condition, we assume that the salt is homogeneously distributed in the domain:
\begin{equation}
    \label{initial_condition}
        c(x,y,z,t=0) = c_0 ~.
\end{equation}
Since $c$ is the only quantity whose time derivative appears in the equations, this initial condition is sufficient for the problem to be well-posed.

\section{\label{Stability_analysis}Linear stability analysis}

\subsection{\label{nondimensional_model}Nondimensional model}

In order to quantitatively investigate the system, it is advantageous to introduce reference values for each physical variable to lower the number of parameters at play and make the equations dimensionless. We choose $D / E$ as the length scale as it relates the advective velocity - in this scenario the evaporation rate - to the diffusivity of salt and therefore is a ratio between the two prevalent transport mechanisms. Furthermore, the reference velocity is chosen as the gravitational velocity $K\rho_0 g c_0 \gamma / \mu$ and the reference time as $\phi D / E^2$. The dimensionless density and salinity are defined such that they are equal to zero at the starting time and track the deviation from the initial state as time goes on. The same approach is used for the pressure, which means that the initial hydrostatic pressure profile has to be subtracted from the actual pressure before the nondimensionalization takes place. Here, both these steps are done at once and the resulting definitions of the nondimensional quantities read:
\begin{equation}
\label{nondim}
        \begin{tabular}{ c c c }
             $\hat{x},\hat{y},\hat{z} = x,y,z~ \dfrac{E}{D}$ , & $\hat{\mathbf{v}} = \mathbf{v} \dfrac{\mu}{K \rho_0 g c_0 \gamma}$ ,  & $\hat{c} = \dfrac{c - c_0}{c_0}$ , \\[0.5cm] $\hat{t} = t \dfrac{E^2}{\phi D}$ , & $\hat{p} = \dfrac{\bigl(p - \rho_0 g z \bigr) E}{D \rho_0 g c_0 \gamma}$ , & $\hat{\rho} = \dfrac{\rho - \rho_0}{\rho_0 c_0 \gamma}$. \\
        \end{tabular}
\end{equation}
From now on, the variables with hat are the dimensionless counterparts of the original variables. Definition \eqref{nondim} yields dimensionless equations in a very convenient form:
\begin{gather}
    \label{nondim_flow_continuity}
    \hat{\nabla} \cdot \hat{\mathbf{v}} = 0~, \\[2pt]
    \label{nondim_darcy}
    \hat{\mathbf{v}} = -\bigl( \hat{\nabla} \hat{p} + \hat{c} ~\mathbf{e_z} \bigr)~, \\[2pt]
    \label{nondim_salt_continuity}
    \partial_{\hat{t}} \hat{c} + \hat{\nabla} \cdot \bigl(\mathrm{Ra} ~ \hat{c} ~\hat{\mathbf{v}} - \hat{\nabla}\hat{c} \bigr) = 0~.
\end{gather}
As $\cn$ and $\hat{\rho}$ are equal to each other, the density term in Darcy's law is simply replaced by the salinity. The Rayleigh number $\mathrm{Ra}$ in this context is defined as 
\begin{equation}
    \mathrm{Ra} = \dfrac{K \rho_0 g c_0 \gamma}{E \mu}~,
\end{equation} 
which can be read as the quotient between the reference velocity as defined in equation \eqref{nondim} and the evaporation rate $E$.
The dimensionless initial and boundary conditions are
\begin{equation}
    \begin{gathered}
    \label{nondim_bc}
    \hat{c}(\hat{x},\hat{y},\hat{z},\tn = 0) = 0~, \\
    \hat{\mathbf{v}}(\hat{x},\hat{y},\zn = 0,\hat{t}) = \hat{\mathbf{v}}(\hat{x},\hat{y},\zn = \alpha,\hat{t}) = \dfrac{1}{\mathrm{Ra}} \mathbf{e_z}~, \\
    \hat{c}(\hat{x},\hat{y},\zn = 0,\hat{t}) = 0 , ~\bigl( \hat{c} +1 - \partial_{\hat{z}} \hat{c} \bigr) \Big|_{\zn = \alpha} = 0~,
    \end{gathered}
\end{equation}
where $\alpha$ is the dimensionless height of the porous slab:
\begin{equation}
    \alpha = \dfrac{H E}{D}~.
\end{equation}
This quantity is an important characteristic of the problem as it quantifies to what degree diffusion is capable of smoothing out the salt concentration over the entire domain height while counteracting advection. As long as there is no convection and the flow velocity is equal to $E$, $\alpha$ also corresponds to the P\'eclet number. Generally, interesting values of $\alpha$ are in the range of $1-1000$ since the diffusion constant of salt in water is of the order $\unitfrac[10^{-9}]{m^2}{s}$ according to Moum et al. \cite{moum1973experimental} and common evaporation rates are roughly $\unitfrac[10^{-8}-10^{-9}]{m}{s}$ depending on the specific ambient conditions \cite{shokri2017impact}. Thus, for heights of $\unit[1-100]{m}$ of the porous layer, we end up with this range for $\alpha$. However, for characteristic heights of 25 or larger and at sufficiently early times, the critical Rayleigh number, which will be introduced in section \ref{eigenvalue_problem}, only differs by 5\% or less from the vertically semi-infinite case that was investigated by Bringedal et al \cite{bringedal2022evaporation}. This will be elucidated in detail in subsection \ref{Ra_over_alpha}. Thus, in this work, we will mainly focus on $\alpha$ values in the range of 1 to 25 and address the effect that varying the dimensionless height has on the critical Rayleigh number.

After the nondimensionalization, we are evidently left with only three pertinent parameters that determine the behaviour of the system : The Rayleigh number, which is of particular significance for buoyancy-driven flows as it determines the occurrence and strength of convection in a fluid layer \cite{kono2001definition}, the characteristic height $\alpha$ and the time $\tn$.

\subsection{\label{ground_state_solution}Ground state}

Before one can look at the occurrence of convection, the stable state, in the following also called ground state, has to be accurately defined. The investigation of perturbations - deviations from the ground state - will then yield stability bounds which determine when convective flow can appear. By definition of the ground-state flow field $\mathbf{\vn}_S$, it has only a nonzero component in the vertical direction which is equal to $1/\mathrm{Ra}$ everywhere due to the boundary conditions \eqref{nondim_bc}. Moreover, the ground-state salinity $\cn_S$ is assumed to only depend on $\zn$. Inserting $\mathbf{\vn}_S$ as flow profile into equation \eqref{salt_conservation} and the boundary conditions \eqref{nondim_bc} yields:
\begin{equation}
    \label{stable_solution_pde}
    \begin{split}
    &\partial_{\tn} \cn_S  + \partial_{\zn} \cn_S - \partial_{\zn}^2 \cn_S = 0~,\\[4pt]
    &\cn_S(\zn,\tn=0) = 0~, \\
    &\cn_S(\zn=0,t) = 0~\quad \cn_S +1 - \partial_{\zn} \cn_S\Big|_{\zn=\alpha} = 0~.
    \end{split}
\end{equation}
These equations describe the time evolution of the salt concentration when water is following a constant upwards flow. By setting $\cn_S(\zn,\tn) = g(\zn,\tn) + \exp{(\zn)} -1$, the system is transformed to a homogeneous boundary value problem for $g$:
\begin{equation}
    \begin{split}
    \label{stable_solution_pde2}
    &\partial_{\tn} g  + \partial_{\zn} g - \partial_{\zn}^2 g = 0~,\\[4pt]
    &g(\zn,0) = 1 - \exp(\zn)~, \\
    &g(\zn = 0,\tn) = 0~\quad g - \partial_{\zn} g \Big|_{\zn=\alpha} = 0~.
    \end{split}
\end{equation}
Such a problem can be solved by separation of variables \cite{polyanin2020functional}, thus we assume $g(\zn,\tn) = X(\zn) \Gamma(\tn)$ with a spatial eigenfunction $X$ and a temporal eigenfunction $\Gamma$. This method leaves us with
\begin{equation}
    \frac{\dot{\Gamma}}{\Gamma} = \frac{X^{''} - X^{'}}{X}~.
\end{equation}
As the left side only depends on $\tn$ and the right side only on $\zn$, both sides have to be equal to some constant $-\lambda$. Now, the equation governing the spatial part of the solution is a Sturm-Liouville problem:
\begin{equation}
    \begin{split}
    \label{Sturm_Liouville_problem}
    &X^{''} - X^{'} = - \lambda X \\
    \Leftrightarrow ~ &\mathcal{L} X := \partial_{\zn} \bigl( \mathrm{e}^{-\zn} \partial_{\zn} \bigr) X = -\lambda \mathrm{e}^{-\zn} X~,
    \end{split}
\end{equation}
where $\mathcal{L}$ is the Sturm-Liouville operator \cite{zettl2010sturm} whose eigenfunctions $X_n$ give us an orthogonal basis for $L^2([0,\alpha],\exp(-\zn))$. Now, the goal is to find the eigenfunctions of $\mathcal{L}$, since this would allow us to write $g(\zn,\tn)$ in terms of these functions with initial coefficients $a_n$, that are determined by the initial condition \eqref{initial_condition}:
\begin{equation}
    \label{separated_solution}
    g(\zn,\tn) = \sum_{n=0}^{\infty} a_n \Gamma_n(\tn) X_n(\zn)~.
\end{equation}
Here, the $\Gamma_n$ are functions capturing the temporal behaviour of every eigenfunction $X_n$. The eigenfunctions of the Sturm-Liouville operator can be found by making an exponential ansatz and solving the characteristic equation. If the coefficients of the resulting function can then be chosen such that the boundary conditions are fulfilled, an eigenfunction is found. Applying this method to the Sturm-Liouville problem \eqref{Sturm_Liouville_problem} together with the boundary conditions of equation \eqref{stable_solution_pde2} leads to the following eigenfunctions:
\begin{equation}
    \begin{split}
    \label{SL_eigenfunctions}
        &X_0(\zn) = \begin{cases}
        \exp{\left(\tfrac{1}{2}\zn\right)} \sin{\left(\delta_n \zn \right)} &\alpha < 2 \\[3pt]
        \zn~ \exp{\left(\frac{1}{2}\zn\right)} &\alpha = 2 \\[3pt]
        \exp{\left(\left(\frac{1}{2}+\epsilon\right)\zn\right)} - \exp{\left(\left(\frac{1}{2}-\epsilon\right)\zn + 2 \epsilon \alpha \right)} \frac{1-2\epsilon}{1+2\epsilon}\quad &\alpha > 2 \end{cases}\\[5pt]
        &X_n(\zn) = \exp{\left(\tfrac{1}{2}\zn\right)} \sin{\left(\delta_n \zn \right)}~ \quad n \in \mathbb{N}^+ ~.
    \end{split}
\end{equation}
The spatial frequencies $\delta_n$ appearing in the eigenfunctions and the value of $\epsilon$ are determined by 
\begin{align}
    \label{delta_n}
    \delta_n - \tfrac{1}{2} \tan{\left(\alpha \delta_n\right)} &= 0~\quad \delta_n>0~, \\[3pt]
    \label{epsilon}
    \exp{\left(2 \epsilon \alpha \right)} \frac{1-2\epsilon}{1+2\epsilon} -1 &=  0~\quad \epsilon>0~.
\end{align}

\begin{figure}[b!]
  \begin{minipage}[c]{0.49\textwidth}
    \centering
    \includegraphics[width=\textwidth]{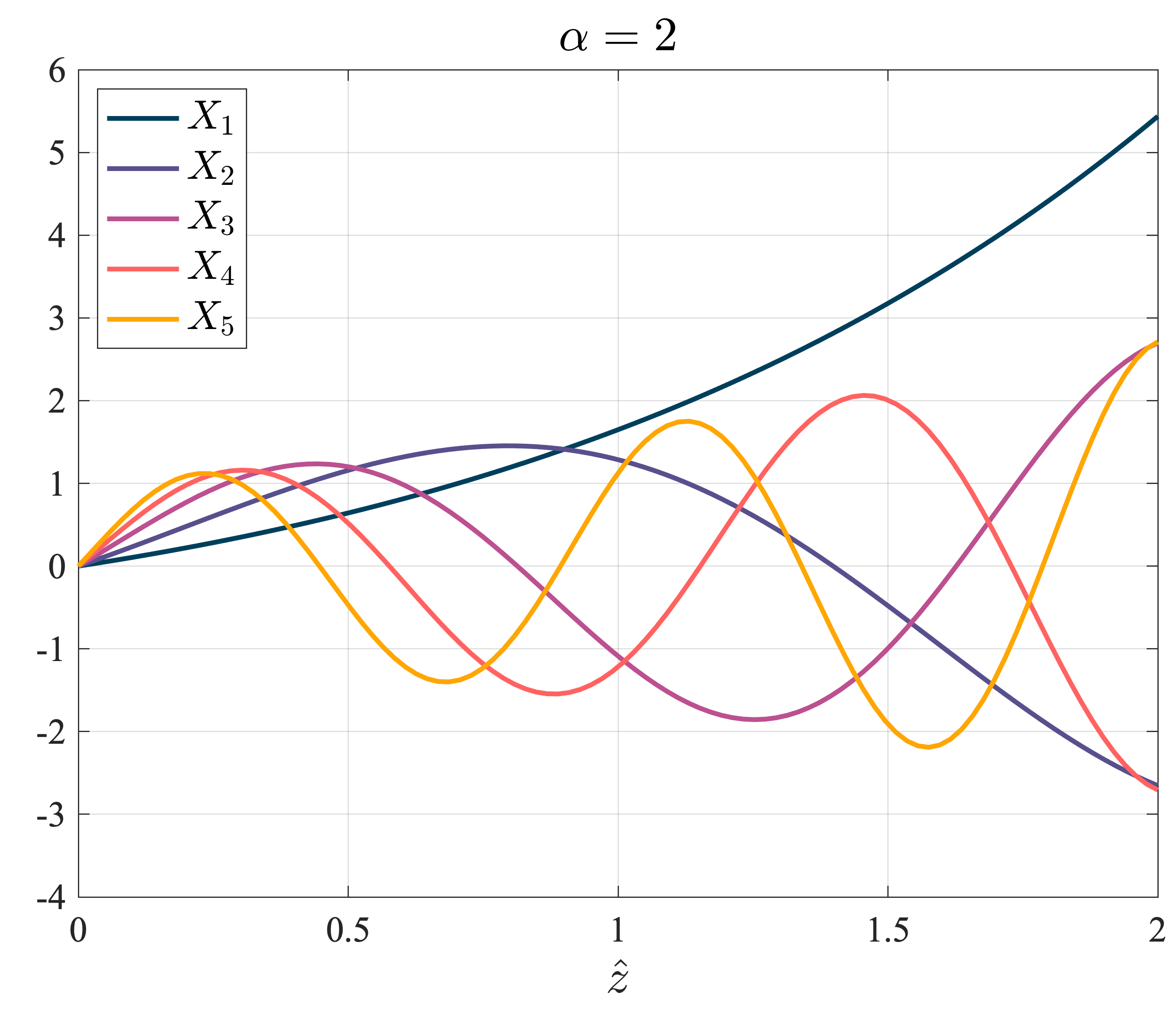}
  \end{minipage}
  \begin{minipage}[c]{0.48\textwidth}
    \centering
    \includegraphics[width=\textwidth]{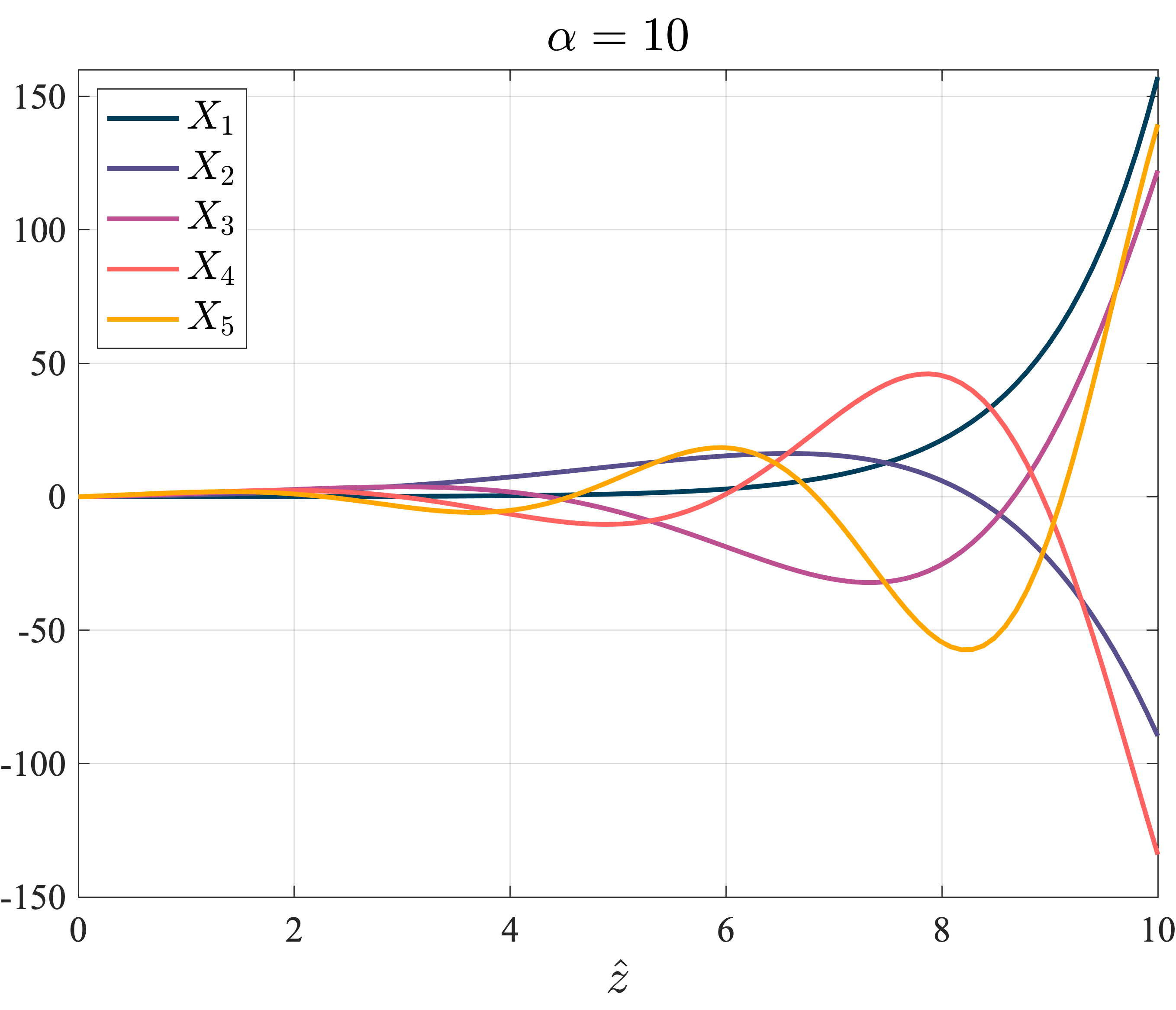}
  \end{minipage}
  \caption[Eigenfunctions of the Sturm-Liouville operator of the ground-state salinity]{The first 5 Sturm-Liouville eigenfunctions as defined in equation \eqref{SL_eigenfunctions} for $\alpha$ equals $2$ and $10$ are shown in the left and right plot, respectively. In the right plot, $X_1$ is scaled by a factor of 1/140 to match the order of magnitude of the other eigenfunctions.}\label{fig:SL_functions}
\end{figure}

Here, equation \eqref{delta_n} is transcendental and has an infinite series of discrete solutions that can be computed numerically. Equation \eqref{epsilon} has only one solution for a given $\alpha > 2$ and has to be numerically solved too. Figure \ref{fig:SL_functions} shows the first 5 Sturm-Liouville eigenfunctions for the exemplary cases of $\alpha$ equals 2 and 10. The eigenvalues corresponding to the eigenfunctions are 
\begin{equation}
    \begin{split}
        &\lambda_0 = \begin{cases}
        \delta_0^2 + \tfrac{1}{4} &\alpha < 2 \\[3pt]
        \tfrac{1}{4} &\alpha = 2 \\[3pt]
        \tfrac{1}{4} - \epsilon^2 &\alpha > 2 \end{cases}~,\\[5pt]
        &\lambda_n = \delta_n^2 + \tfrac{1}{4}~ \quad n \in \mathbb{N}^+ ~.
    \end{split}
\end{equation}

Now, we have assembled everything needed to compute the ground-state salt concentration. Inserting the approach from equation \eqref{separated_solution} into the differential equation \eqref{stable_solution_pde} and solving for the time functions $\Gamma_n$ gives
\begin{equation}
    \Gamma_n(\tn) = \exp{\left(-\lambda_n \tn \right)}~.
\end{equation}
The coefficients $a_n$ for the transformed initial condition in equation \eqref{stable_solution_pde2} are determined by 
\begin{equation}
    \label{SL_initial_coefficients}
    a_n = \ddfrac{\int_0^\alpha \bigl(1-\exp(\zn)\bigr) X_n(\zn) \exp(-\zn) \mathrm{d}\zn}{\int_0^\alpha X_n^2(\zn) \exp(-\zn) \mathrm{d}\zn}~,
\end{equation}

\begin{figure*}[t!]
  \begin{minipage}[c]{0.49\textwidth}
    \centering
    \includegraphics[width=\textwidth]{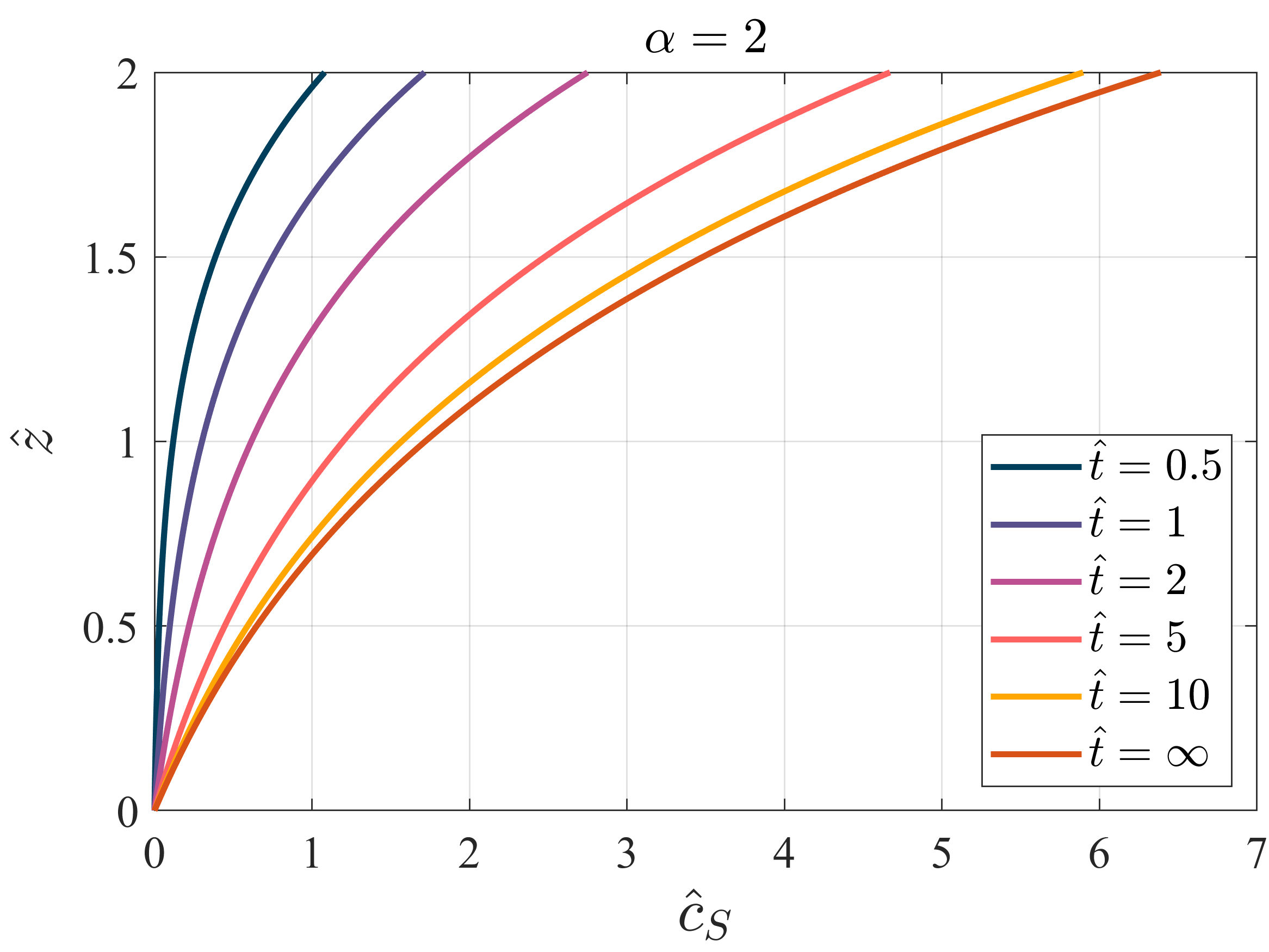}
  \end{minipage}
  \begin{minipage}[c]{0.49\textwidth}
    \centering
    \includegraphics[width=\textwidth]{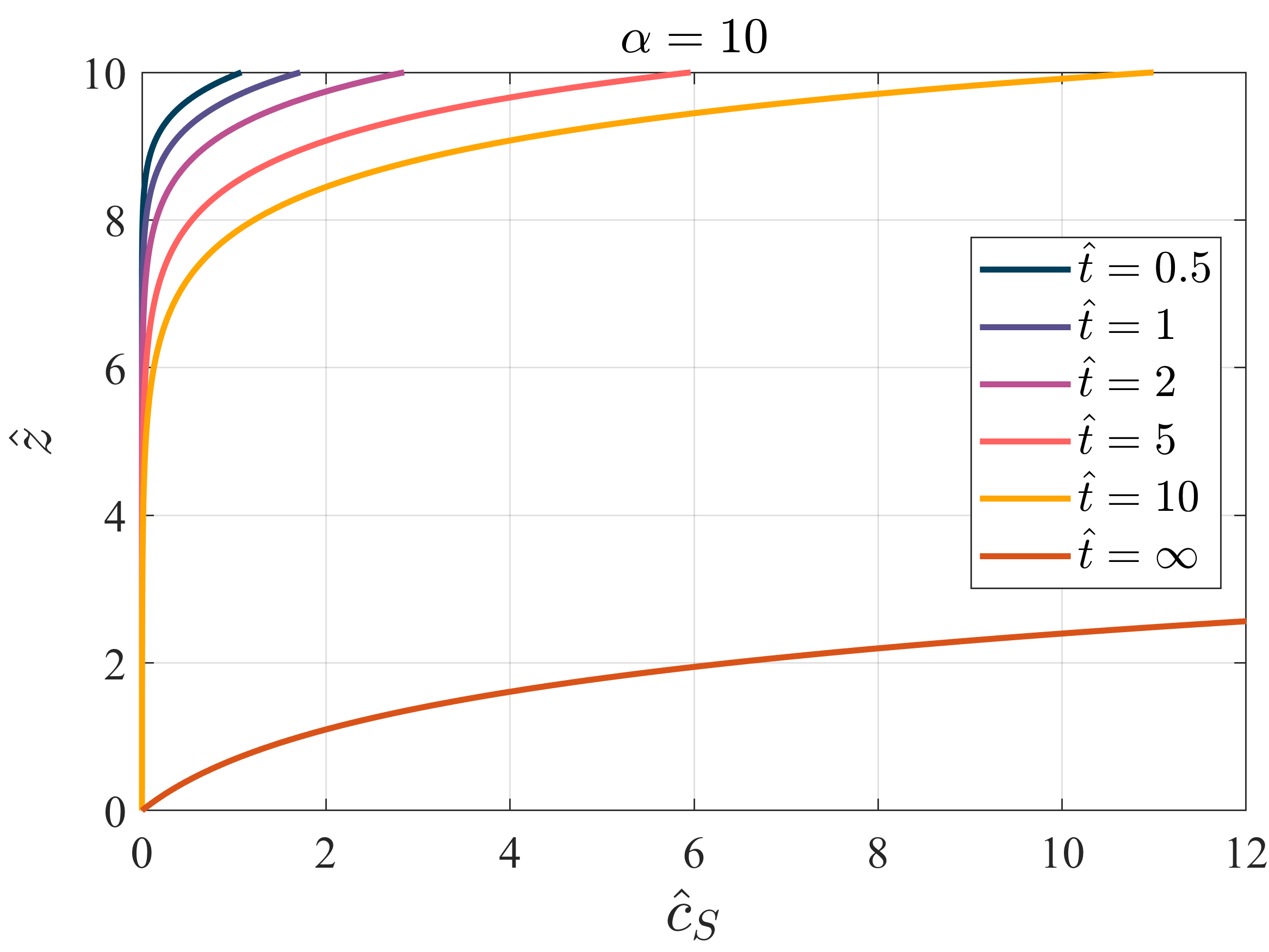}
  \end{minipage}
  \caption[Ground-state salinity for $\alpha$ equals 2 and 10 at different times]{The ground-state salt concentration is plotted at different times. It is visible that the ground state converges much faster towards the steady state in the smaller domain, whereas for $\alpha$ equals $10$ the salt has accumulated at the top of the domain but has yet to diffuse to the bottom at $\tn = 10$.}\label{fig:c_S}
\end{figure*}

since the Sturm-Liouville eigenfunctions are orthogonal with respect to the $L^2$ inner product with weight function $\exp(-\zn)$ \cite{zettl2010sturm}. Putting everything together gives us the ground-state velocity, salinity and pressure:
\begin{align}
    \label{v_S}
    \mathbf{\vn}_S(\zn, \tn) &= \frac{1}{\mathrm{Ra}} \mathbf{e_z}~,\\[2pt]
    \label{c_S}
    \cn_S(\zn,\tn) &= \exp(\zn) -1 + \sum_{n=0}^{\infty} a_n \Gamma_n(\tn) X_n(\zn)~,\\
    \label{p_S}
    \pn_S(\zn, \tn) &= -\frac{\zn}{\mathrm{Ra}} - \int_0^{\zn} \cn_S(\xi,\tn) \mathrm{d}\xi + C_{\pn}(\tn)~,
\end{align}
where the pressure is derived by integrating Darcy's law and the constant $C_{\pn}(\tn)$ can be chosen arbitrarily. Evidently, the salt concentration converges against the offsetted exponential function, since all eigenvalues $\lambda_n$ are positive and, thus, all the terms in the infinite sum decay over time. Hence, the steady-state solution is given by:
\begin{align}
    \label{v_SS}
    \mathbf{\vn}_S(\zn, \tn = \infty) &= \frac{1}{\mathrm{Ra}} \mathbf{e_z}~,\\[2pt]
    \label{c_SS}
    \cn_S(\zn,\tn = \infty) &= \exp(\zn) -1~,\\
    \label{p_SS}
    \pn_S(\zn, \tn = \infty) &= \zn \left( 1 - \frac{1}{\mathrm{Ra}}\right) - \exp(\zn) + 1 + C_{\pn}(\tn)~.
\end{align}

In figure \ref{fig:c_S} we can see $\cn_S$ for $\alpha$ equal to 2 and 10 at different times. As expected, the salt concentration increases at the top due to the zero-flow Robin boundary condition and is transported downwards with time due to diffusion. It is also visible that $\cn_S$ converges much faster against its steady state for small $\alpha$. From a physical point of view, the steady state is reached when the diffusion of salt out of the domain at the bottom is equal to the advective inflow. For large $\alpha$, a longer period of time is needed to get to this stage as the diffusive transport from the top boundary has to cover a bigger distance.

\begin{figure}[t!]
\floatbox[{\capbeside\thisfloatsetup{capbesideposition={right,top},capbesidewidth=5.8cm}}]{figure}[\FBwidth]
{\caption[Comparison of ground-state salt concentrations for different $\alpha$]{The plot shows $\cn_{S}(\zn,\tn)$ for $\alpha$ equal to 5 and 50 at different times. The ground-state salt concentrations are plotted at a vertical distance of 0 to 5 from the top for both $\alpha$.} \label{fig:ground_state_comparison}}
{\includegraphics[width=6cm]{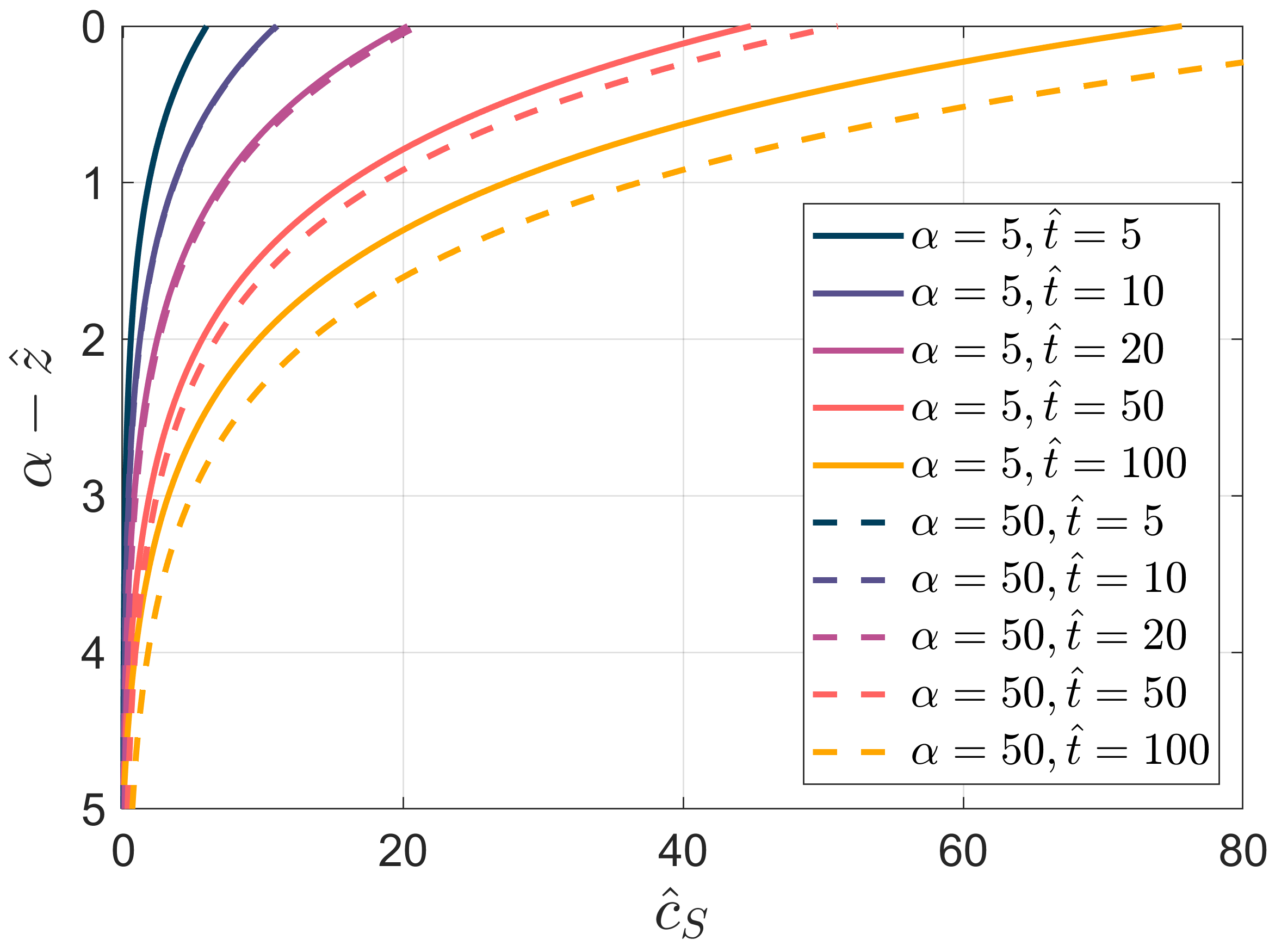}}
\end{figure}

At early times, such that the salt accumulation at the top has not been distributed through the entire domain by diffusion - meaning only very small amounts of salt leave the slab through the bottom boundary - the solutions for different $\alpha$ have a very similar shape. This is visible in figure \ref{fig:ground_state_comparison}, where the ground states for $\alpha$ equal to 5 and 50 are plotted in the interval $[\alpha -5, \alpha]$. In the case of $\alpha$ being equal to 5, this interval corresponds to $\zn \in [0,5]$, whereas for $\alpha$ equal to 50, the ground-state solution is plotted over the interval $\zn \in [45,50]$. As visible in figure \ref{fig:ground_state_comparison}, both ground-state salt concentrations only begin to significantly deviate from each other at $\tn$ equal to 100. This resemblance of the ground states at sufficiently small times is also reflected in the similarity of the critical Rayleigh number for differing characteristic heights as we will see in section \ref{Ra_over_alpha}.

\subsection{\label{linear_perturbation_equations}Linear perturbation equations }

Our goal is to investigate the stability of the ground state as defined in equations \eqref{v_S}, \eqref{c_S}, \eqref{p_S}. The method of linearized stability will be employed to calculate the critical Rayleigh number. Therefore, equations governing the behaviour of infinitesimal perturbations to the ground state have to be derived. We can study the behaviour of perturbations to the ground-state solution by assuming that the actual solution is of the form $\mathbf{\vn} = \mathbf{\vn}_S + \mathbf{\vn}'$, $\cn = \cn_S + \cn'$ and $\pn = \pn_S + \pn'$ where $\mathbf{\vn}', \pn', \cn'$ are small perturbations. Inserting this first-order perturbation approach in the original equations \eqref{nondim_flow_continuity}, \eqref{nondim_darcy}, \eqref{nondim_salt_continuity} gives us a new system of perturbation equations. Now, by only considering infinitesimal perturbations, we can neglect second- or higher-order terms which yields the linearized equations
\begin{gather}
     \label{perturb_flow_continuity}
    \nabla \cdot \mathbf{\vn}' = 0~, \\[2pt]
    \label{perturb_darcy}
    \mathbf{\vn}' = -\bigl( \nabla \pn' + \cn' \mathbf{e_z} \bigr)~, \\[2pt]
    \label{perturb_salt_continuity}
    \partial_{t} \cn' + \mathrm{Ra}~ \partial_{\zn} \cn_S~ \mathrm{\vn}'_{\zn} - \partial_{\zn} \cn' = \nabla^2 \cn'~,
\end{gather}
and leads to the homogeneous boundary conditions for the perturbed quantities:
\begin{equation}
\label{perturb_boundary_conditions}
    \begin{gathered}
        \mathbf{\vn}'(\xn,\yn,\zn=0,\tn) = \mathbf{\vn}'(\xn,\yn,\zn=\alpha,\tn) = 0~, \\
        \cn'(\xn,\yn,\zn=0,\tn) = 0 , ~ \cn' - \partial_{\zn} \cn' \Big|_{\zn=\alpha} = 0~.
    \end{gathered}
\end{equation}
The solenoidal vector field $\vn'$ is fully determined by the vertical flow strength $\omega := \mathbf{\vn}'_z$ and the vertical component $\zeta$ of the vorticity \cite{homsy1976convective}. By taking the curl of equation \eqref{perturb_darcy}, one can show that $\zeta$ is equal to zero. Consequently, $\omega$ and $\chi := \cn'$ are the only scalar quantities that characterize the perturbation system. Taking the vertical component of the double curl of equation \eqref{perturb_darcy} leads to 
\begin{equation}
    \label{perturbation_eq2}
    \nabla^2 \omega = -\bigl( \partial_{\xn}^2 + \partial_{\yn}^2 \bigr) \chi~.
\end{equation} 
This equation together with \eqref{perturb_salt_continuity} are Fourier decomposable in $\xn$ and $\yn$ direction. Consequently, we will consider perturbations of the form
\begin{equation}
    \begin{split}
    \label{perturbation_ansatz}
    \omega(\xn, \yn, \zn, \tn) = \omega(\zn) \mathrm{e}^{\sigma \tn} \cos(\an_x \xn) \cos(\an_y \yn)~, \\
    \chi(\xn,\yn,\zn,\tn) = \chi(\zn) \mathrm{e}^{\sigma \tn} \cos(\an_x \xn) \cos(\an_y \yn)~,
    \end{split}
\end{equation}
where $\omega$ and $\chi$ are the vertical perturbation eigenfunctions that only depend on $\zn$. Here, $\sigma$ is the growth rate of the perturbation and $\an_x$ and $\an_y$ are the wavenumbers of the perturbation in the lateral directions. This is a standard approach when examining linear perturbation equations where the coefficients do not depend on time and horizontal coordinates $\xn,\yn$ \cite{chan2012linear}. Since the ground-state salt concentration $\cn_S(\zn, \tn)$ depends on time, the separation of variables in equation \eqref{perturbation_ansatz} does not directly apply. However, one can assume that the ground state changes slower in time than any exponentially growing instability which allows for a separation of ground-state time $\tn$ and perturbation time $\tau$, commonly known as frozen profile approach or quasi-steady state approximation \cite{chan2012linear}. Furthermore, van Duijn et al. \cite{van2001stability} have shown that if the Rayleigh number $\mathrm{Ra}$ of a physical system is strictly bigger than the smallest eigenvalue of the perturbation equations for a growth rate $\sigma$ of zero, then there exists a growing infinitesimal perturbation  which implies that the boundary layer is unstable. Thus, in order to find a stability bound, it is sufficient to analyze perturbations only for the case of neutral stability. Setting $\sigma$ to zero and inserting ansatz \eqref{perturbation_ansatz} in equation \eqref{perturb_salt_continuity} and \eqref{perturbation_eq2} while assuming the quasi-steady state approximation, gives us the final perturbation equations:
\begin{gather}
    \label{perturb_ode_w}
    \bigl( \partial_{\zn}^2 - \an^2 \bigr) \omega - \an^2 \chi = 0~, \\
    \label{perturb_ode_c}
    \bigl( \partial_{\zn}^2 - \partial_{\zn} - \an^2 \bigr) \chi - \mathrm{Ra}~ \partial_{\zn} \cn_S(\zn, \tn) ~ \omega = 0~,
\end{gather}
where $\an := \sqrt{\hat{a}_x^2 + \hat{a}_y^2}$ is the wavenumber of the perturbations in the $\xn,\yn$ plane. Note that the time $\tn$ only appears as a parameter in equations \eqref{perturb_ode_w}, \eqref{perturb_ode_c}, which will allow us to obtain a stability bound at every time of the ground state.

\subsection{\label{eigenvalue_problem} The eigenvalue problem}

The two equations \eqref{perturb_ode_w}, \eqref{perturb_ode_c} with the boundary conditions \eqref{perturb_boundary_conditions} govern the shape of infinitesimal, admissible perturbations to the ground state. Due to the homogeneity of the equations, setting $\omega$ and $\chi$ to zero is a valid solution of the perturbation boundary value problem. Hence, instabilities in the linear limit can only develop when the perturbation equations have multiple solutions and we can derive a stability criterion by investigating the uniqueness of solutions of the perturbation system. The resulting eigenvalue problem reads:

Given $\alpha > 0$, $\tn > 0$  and $\an > 0$, find the smallest $\mathrm{Ra} = \mathrm{Ra}(\alpha,\tn,\an) > 0$, such that
\begin{equation}
    \begin{cases}
        \bigl( \partial_{\zn}^2 - \an^2 \bigr) \omega - \an^2 \chi = 0 \quad &\zn \in (0,\alpha)\\
        \bigl( \partial_{\zn}^2 - \partial_{\zn} - \an^2 \bigr) \chi - \mathrm{Ra}~ \partial_{\zn} \cn_S(\zn, \tn) \omega = 0 \quad &\zn \in (0,\alpha) \\
        \omega = 0, ~ \chi = 0 \quad &\zn = 0\\
        \omega = 0, ~ \chi -\partial_{\zn} \chi = 0 \quad &\zn = \alpha
    \end{cases}~,
\end{equation}
has non-trivial solutions, where $\cn_S(\zn, \tn)$ is defined as in equation \eqref{c_S}. 

For a given $\alpha$ and $\tn$, the critical Rayleigh number $\mathrm{Ra_c}$ is now defined as the smallest eigenvalue for all possible wavelengths. Since this work considers a laterally unbounded domain, the wavelengths can be arbitrary:
\begin{equation}
    \label{critical_rayleigh}
    \mathrm{Ra_c}(\alpha, \tn) = \min_{\an > 0} \mathrm{Ra}(\alpha, \tn, \an)~.
\end{equation}
The critical Rayleigh number has an important physical meaning as it separates the unstable from the stable regime. If an actual system with dimensionless height $\alpha_s$ at time $\tn_s$ has a Rayleigh number $\mathrm{Ra_s} > \mathrm{Ra_c}(\alpha_s, \tn_s)$, an exponentially growing perturbation exists, corresponding to convective instabilities.

\subsection{\label{solution_methods} Solution strategies for the eigenvalue problem}

\subsubsection{\label{fundamentalmatrix_method}Fundamental matrix method}

One approach to scrutinize whether a solution of a linear boundary-value problem is unique is to rewrite the corresponding ordinary differential equations into a system of first-order ordinary differential equations (ODE). We denote the linear system of ODEs with:
\begin{equation}
    \label{def_linear_ode_system}
    \partial_\xi \mathbf{u} = \mathbf{A}(\xi) \mathbf{u} + \mathbf{b}(\xi) ~,\quad \xi \in \mathbb{I} \subseteq \mathbb{R}
\end{equation}
where $\mathbf{A} \in C(\mathbb{I};\mathbb{R}^{d\times d})$ and  $\mathbf{b} \in C(\mathbb{I};\mathbb{R}^{d})$. The associated linear boundary-value problem is then denoted as
\begin{equation}
    \label{def_linear_boundary_condition}
    \mathbf{R_1}\mathbf{u}(\xi_0) +\mathbf{R_2}\mathbf{u}(\xi_1) - \mathbf{r_0} = 0 ~, \quad \xi_0,\xi_1 \in \mathbb{I}
\end{equation}
with appropriate matrices $\mathbf{R_1}, \mathbf{R_2} \in \mathbb{R}^{d \times d}$ and a vector $\mathbf{r_0} \in \mathbb{R}^d$ to cover inhomogeneous boundary conditions. Here, $\xi_0$ and $\xi_1$ are the points in time or locations in space where certain boundary conditions are imposed on the unknown $\mathbf{u}$. For such systems, the fundamental matrix $\Phi(\cdot;\xi_0) \in C^1(\mathbb{I},\mathbb{R}^{d\times d})$ is of central importance. This matrix is well-defined by the following properties \cite{somasundaram2001ordinary}:
\begin{equation}
    \label{def_fundamentalmatrix}
    \begin{split}
        \partial_\xi \Phi(\xi;\xi_0) &= \mathbf{A}(\xi) \Phi(\xi;\xi_0)~, \\
        \Phi(\xi_0;\xi_0) &= \mathbf{I_d}~.
    \end{split}
\end{equation}
Here, $\mathbf{I_d}$ is the d-dimensional identity matrix. Hence, the fundamental matrix is a matrix-valued function whose columns are the linearly independent solutions $\mathbf{u_i}(\xi)$ of the homogeneous, linear ODE system for the initial condition $\mathbf{u_i}(\xi_0) = \mathbf{e_i}$, where $\mathbf{e_i}$ is the i-th basis vector.

Having now defined all relevant quantities, the solution of the boundary-value problem consisting of differential equation system \eqref{def_linear_ode_system} and linear boundary conditions defined in equation \eqref{def_linear_boundary_condition} is unique if 
\begin{equation}
    \label{uniqueness-theorem}
    \det{\bigl( \mathbf{M}(\xi_0) \bigr)} := \det{\bigl( \mathbf{R_1} + \mathbf{R_2} \Phi(\xi_1;\xi_0)\bigr)} \neq 0~.
\end{equation}
The idea of this new approach is now to leverage this statement to solve the eigenvalue problem in section \ref{eigenvalue_problem}. Therefore, the perturbation equations \eqref{perturb_ode_w} and \eqref{perturb_ode_c} have to be rewritten into a 4-dimensional, first-order ODE system. That approach leads to
\begin{equation}
\label{4d_ode}
\partial_{\zn} \left(
    \begin{matrix}
         u_1 \\
         u_2 \\
         u_3 \\
         u_4 \\
    \end{matrix}
    \right)
    = 
    \left(
    \begin{matrix}
        0 & 1 & 0 & 0 \\
        \an^2 & 0 & \an^2 & 0 \\
        0 & 0 & 0 & 1 \\
        \mathrm{Ra}~ \partial_{\zn} \cn_S & 0 & \an^2 & 1
    \end{matrix}
    \right)
    \left(
    \begin{matrix}
        u_1 \\
        u_2 \\
        u_3 \\
        u_4 \\
    \end{matrix}
    \right) =: \mathbf{A}(\zn) ~\mathbf{u}~,
\end{equation}
in our case, where $u_1 = \omega$, $u_2 = \partial_{\zn} \omega$, $u_3 = \chi$ and $u_4 = \partial_{\zn} \chi$. The linear boundary conditions \eqref{perturb_boundary_conditions} can be written in the form
\begin{equation}
    \label{linear_bc}
    \left(
    \begin{matrix}
        1 & 0 & 0 & 0 \\
        0 & 0 & 0 & 0 \\
        0 & 0 & 1 & 0 \\
        0 & 0 & 0 & 0
    \end{matrix}
    \right)
    \left(
    \begin{matrix}
        u_1(0) \\
        u_2(0) \\
        u_3(0) \\
        u_4(0) \\
    \end{matrix}
    \right) + 
    \left(
    \begin{matrix}
        0 & 0 & 0 & 0 \\
        1 & 0 & 0 & 0 \\
        0 & 0 & 0 & 0 \\
        0 & 0 & 1 & -1
    \end{matrix}
    \right)
    \left(
    \begin{matrix}
        u_1(\alpha) \\
        u_2(\alpha) \\
        u_3(\alpha) \\
        u_4(\alpha) \\
    \end{matrix}
    \right) =: \mathbf{R_1} \mathbf{u}(0) + \mathbf{R_2}\mathbf{u}(\alpha) = 0~.
\end{equation}
Now, the fundamental matrix of system \eqref{4d_ode} has to be computed before equation \eqref{uniqueness-theorem} can be employed. Note that for system matrices $\mathbf{A}$ that commutate with their own elementwise integral, one can calculate the fundamental matrix by \cite{somasundaram2001ordinary}
\begin{equation}
    \label{fundamentalmatrix}
    \Phi(\xi;\xi_0) = \exp{\left(\int_{\xi_0}^{\xi} \mathbf{A}(\nu) \mathrm{d}\nu \right)}~.
\end{equation}
However, this is not the case for $\mathbf{A}$ as defined in equation \eqref{4d_ode}, due to the appearance of $\partial_{\zn}\cn_S$ as a $\zn$-dependent term in its forth row. Hence, for our purposes another method is required to calculate $\Phi(\alpha; 0)$, which is the required quantity to evaluate the uniqueness of the solution of the perturbation equations by virtue of equation \eqref{uniqueness-theorem}.

As stated in the work of Balser et al. \cite{balser2006systems}, when the system matrix $\mathbf{A}$ of a linear ODE system is analytic on some open interval around the initial point $\xi_0$, then it can be expanded into its power series and there also exists a power series solution for the fundamental matrix. Since the only non-constant term in $\mathbf{A}$ in our case is the derivative of the ground state salinity $\partial_{\zn} \cn_S$, which is analytic, $\mathbf{A}$ is analytic on the entire interval $[0,\alpha]$. Consequently, we can expand $\mathbf{A}$ and $\Phi$ into a power series
\begin{equation}
    \mathbf{A}(\zn) = \sum_{k=0}^{\infty} (\zn-\zn_0)^k ~ \mathbf{A}_k~, \quad \zn \in \mathbb{R}
\end{equation}
around $\zn_0$, which will be 0 in our case, since this is the height where the bottom boundary conditions are imposed. Thus, the coefficient matrices $\mathbf{A_k}$ are equal to
\begin{equation}
\label{system_matrix_coefficients}
    \mathbf{A}_0 = 
    \left(\begin{matrix}
        0 & 1 & 0 & 0 \\
        \an^2 & 0 & \an^2 & 0\\
        0 & 0 & 0 & 1\\
        \mathrm{Ra} ~\partial_{\zn} \cn_S (0) & 0 & \an^2 & 1\\
    \end{matrix}\right),\quad 
    \mathbf{A}_i = 
    \left(\begin{matrix}
        0 & 0 & 0 & 0 \\
        0 & 0 & 0 & 0\\
        0 & 0 & 0 & 0\\
        \frac{\mathrm{Ra}}{i!} ~\partial_{\zn}^{(i+1)} \cn_S (0) & 0 & 0 & 0\\
    \end{matrix}\right) ~~~ i \in \mathbb{N}^+ ~,\\
\end{equation}
by considering the Taylor expansion of $\partial_{\zn} \cn_S$ around 0. The derivatives of order $i+1$ appearing in $\mathbf{A}_i$ are computed via a recursion relation of the derivative of the Sturm-Liouville eigenfunctions appearing in the ground-state salinity. Following Balser et al.\cite{balser2006systems}, expanding $\boldsymbol{\Phi}$ in its power series is now also possible:
\begin{equation}
    \label{fundamentalmatrix_power_series}
    \Phi(\zn;\zn_0) = \sum_{k=0}^{\infty} (\zn-\zn_0)^k ~\Phi_k~.
\end{equation}
The coefficients matrices $\Phi_k$ of the fundamental matrix can be determined by comparison of coefficients and, thus, have to be equal to
\begin{align}
        \Phi_{k+1} &= \dfrac{1}{k+1} \sum_{j=0}^{k} ~ \mathbf{A}_{k-j} \Phi_j~, \\
        \Phi_0 &= \mathbb{I}~.
\end{align}
The power series expansion of the fundamental matrix in equation \eqref{fundamentalmatrix_power_series} can be evaluated at $\zn = \alpha$ in order to compute $\Phi(\alpha;0)$. The roots of the determinant in equation \eqref{uniqueness-theorem} are found by defining a fixed grid of $\alpha$ and $\an$ values at each time $\tn$ and treating $\mathrm{Ra}$ as the only unknown at every grid point. The one-dimensional root search to find the corresponding eigenvalue is done via the bisection method and the critical Rayleigh number $\mathrm{Ra}_c(\alpha, \tn)$ is calculated by finding the smallest Rayleigh number for all wavenumbers according to equation \eqref{critical_rayleigh}. 

This new method only relies on the convergence of the power series of the system matrix that describes the first-order ODE system corresponding to the perturbation equations. Hence, it can be applied to a wide range of perturbation eigenvalue problems such as the Orr-Sommerfeld equation or Rayleigh's equation \cite{craik1988wave}.

\subsubsection{\label{chebyshev_galerkin_method} Chebyshev-Galerkin method}

Due to deficiencies of the fundamental matrix method under certain conditions, which will be elucidated in section \ref{stability_limit}, we also introduce and employ a Petrov Chebyshev-Galerkin method to solve the eigenvalue problem emerging from the perturbation equations. Therefore, we set
\begin{equation}
    \begin{split}
    \label{chebyshev_ansatz}
        \omega(\zn) = \sum_{n=0}^{\infty} \omega_n T_n(\zn), \quad \chi(\zn) = \sum_{n=0}^{\infty} \chi_n T_n(\zn),
    \end{split}
\end{equation}
where $T_n$ are the affinely transformed Chebyshev polynomials
\begin{equation}
    T_n: [0, \alpha] \rightarrow [-1,1], \quad ~T_n(\zn) = \tilde{T}_n(2\dfrac{\zn}{\alpha}-1)~,
\end{equation}
and $\tilde{T}_n$ are the Chebyshev polynomials of the first kind \cite{mason2002chebyshev}. Inserting the ansatz \eqref{chebyshev_ansatz} into the perturbation equations \eqref{perturb_ode_w}, \eqref{perturb_ode_c}, multiplying with $T_m$ and integrating over $\zn$ yields
\begin{equation}
    \begin{split}
    \label{chebyshev_galerkin_equations}
        &\sum_{n=0}^{\infty} \omega_n \Bigl( \int_0^\alpha T_n^{''} T_m ~\mathrm{d}\zn - \an^2 \int_0^\alpha T_n T_m ~\mathrm{d}\zn \Bigr) \\ &\quad - \sum_{n=0}^{\infty} \chi_n \an^2 \int_0^\alpha T_n T_m ~\mathrm{d}\zn = 0~, \\
        &\sum_{n=0}^\infty \chi_n \Bigl(\int_0^\alpha T_n^{''} T_m \mathrm{d}\zn - \int_0^\alpha T_n^{'} T_m \mathrm{d}\zn - \an^2 \int_0^\alpha T_n T_m \mathrm{d}\zn \Bigr) \\ &\quad - \sum_{n=0}^\infty \omega_n \mathrm{Ra} \int_0^\alpha T_n T_m \partial_{\zn} \cn_S \mathrm{d}\zn = 0~.
    \end{split}
\end{equation}

In order to get a finite system of equations, the sums in equation \eqref{chebyshev_galerkin_equations} are truncated after the first N terms and only the first N-2 Chebyshev polynomials are used as test functions for each equation. Since the boundary conditions for $\omega$ and $\chi$ are not built into the ansatz space, they also have to be enforced and yield 4 additional equations:
\begin{gather}
\begin{aligned}
    \label{galerkin_boundary_conditions}
        \omega(0) &= \sum_{n=0}^N \omega_n (-1)^n = 0~, \quad &\omega(\alpha) = \sum_{n=0}^N \omega_n = 0~, \\
        \chi(0) &= \sum_{n=0}^N \chi_n (-1)^n = 0~,\quad &\chi(\alpha) - \partial_{\zn}\chi(\alpha) = \sum_{n=0}^N \chi_n \biggl(1-\frac{2}{\alpha}k^2\biggr) = 0~.
\end{aligned}
\end{gather}
Usually, the boundary conditions are integrated into the ansatz space via basis recombination. This is also possible in this scenario. However, combining Chebyshev polynomials such that the Robin boundary condition of the salinity is fulfilled leads to convoluted expressions. Since the results of enforcing the boundary conditions with additional equations are similar to building and employing a recombined basis, the former approach is taken for the sake of simplicity.

Equations \eqref{galerkin_boundary_conditions} together with the $2(N-2)$ equations in \eqref{chebyshev_galerkin_equations} constitute a linear equation system $\mathbf{M}~ \mathbf{b} = \mathbf{0}$ determined by a quadratic block matrix
\begin{equation}
\label{matrix_chebyshev}
\begin{gathered}
    \mathbf{M}(\alpha,\an,\mathrm{Ra},\tn)~ =\left(
\begin{array}{cc}
\mathbf{U}(\alpha,\an) & \mathbf{V}(\alpha,\an) \\
\mathrm{Ra}\cdot \mathbf{W}(\alpha,\an,\tn) & \mathbf{X}(\alpha,\an)
\end{array}\right)  \in \mathbb{R}^{N \times N},
\end{gathered}
\end{equation}
where the last two rows of $\mathbf{U}$, $\mathbf{V}$ and $\mathbf{W}$, $\mathbf{X}$ contain the boundary conditions for $\omega$ and $\chi$, respectively. The eigenvalues of the perturbation equations can now again be found by searching for the roots of the determinant of $\mathbf{M}$, since otherwise the equations system has only the zero solution. Using a formula for the determinant of a block matrix \cite{silvester2000determinants}, the root search of the determinant can be rewritten into a generalized eigenvalue problem:
\begin{equation}
    \label{generalized_eigenproblem}
    \mathbf{U} ~\Lambda =  \mathrm{Ra}~\left(\mathbf{V} \mathbf{X}^{-1} \mathbf{W}\right)~ \Lambda, \quad \Lambda \in \mathbb{R}^M~.
\end{equation}
The benefit of this reformulation is that there are more efficient solving methods available. According to the definition of the perturbation eigenvalue problem in section \ref{eigenvalue_problem}, calculating the smallest eigenvalue of the generalized eigenvalue problem will yield $\mathrm{Ra}(\alpha, \tn, \an)$.

\subsection{\label{stability_limit} The stability limit}

Both the fundamental matrix method and the Chebyshev-Galerkin method are employed to solve the eigenvalue problem. For all the following results, the series expansion of the fundamental matrix is truncated after the first 150 terms, due to the convergence of the calculated eigenvalues for $\alpha$ between 1 and 25. The first 30 Chebyshev polynomials are used as ansatz functions for the perturbed vertical velocity and salt concentration in the Chebyshev-Galerkin method. The series of Sturm-Liouville eigenfunctions in the ground state salt concentration as defined in equation \eqref{c_S} is truncated after the first 100 terms since the maximal difference on the whole domain between using 100 and 500 terms to calculate the ground-state salinity is of the order of $10^{-10}$ for $\alpha \in [1,25]$.

\subsubsection{\label{Ra_over_a} Rayleigh number as a function of the wavenumber}

\begin{figure*}[t!]
  \caption[Rayleigh number over wavenumber $\an$]{The plot shows the eigenvalues $\mathrm{Ra}(\alpha, \tn, \an)$ of the perturbation system as a function of $\an$ for $\alpha = 2$ and 5, respectively, at varying times between 0.5 and infinity.}
  \label{fig:Ra_over_a}
  {\begin{minipage}[c]{0.49\textwidth}
    \centering
    \includegraphics[width=\textwidth]{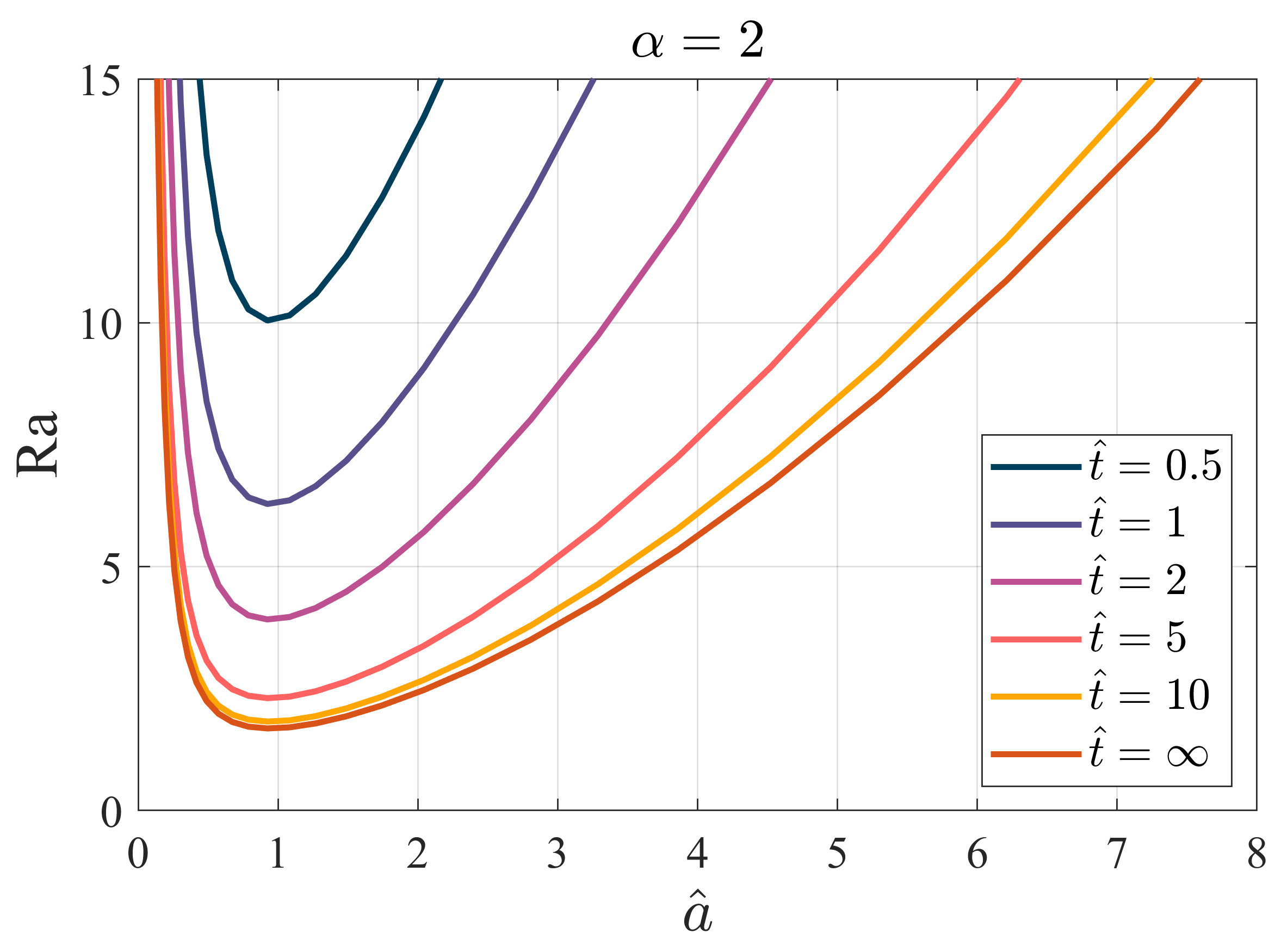}
  \end{minipage}
  \begin{minipage}[c]{0.49\textwidth}
    \centering
    \includegraphics[width=\textwidth]{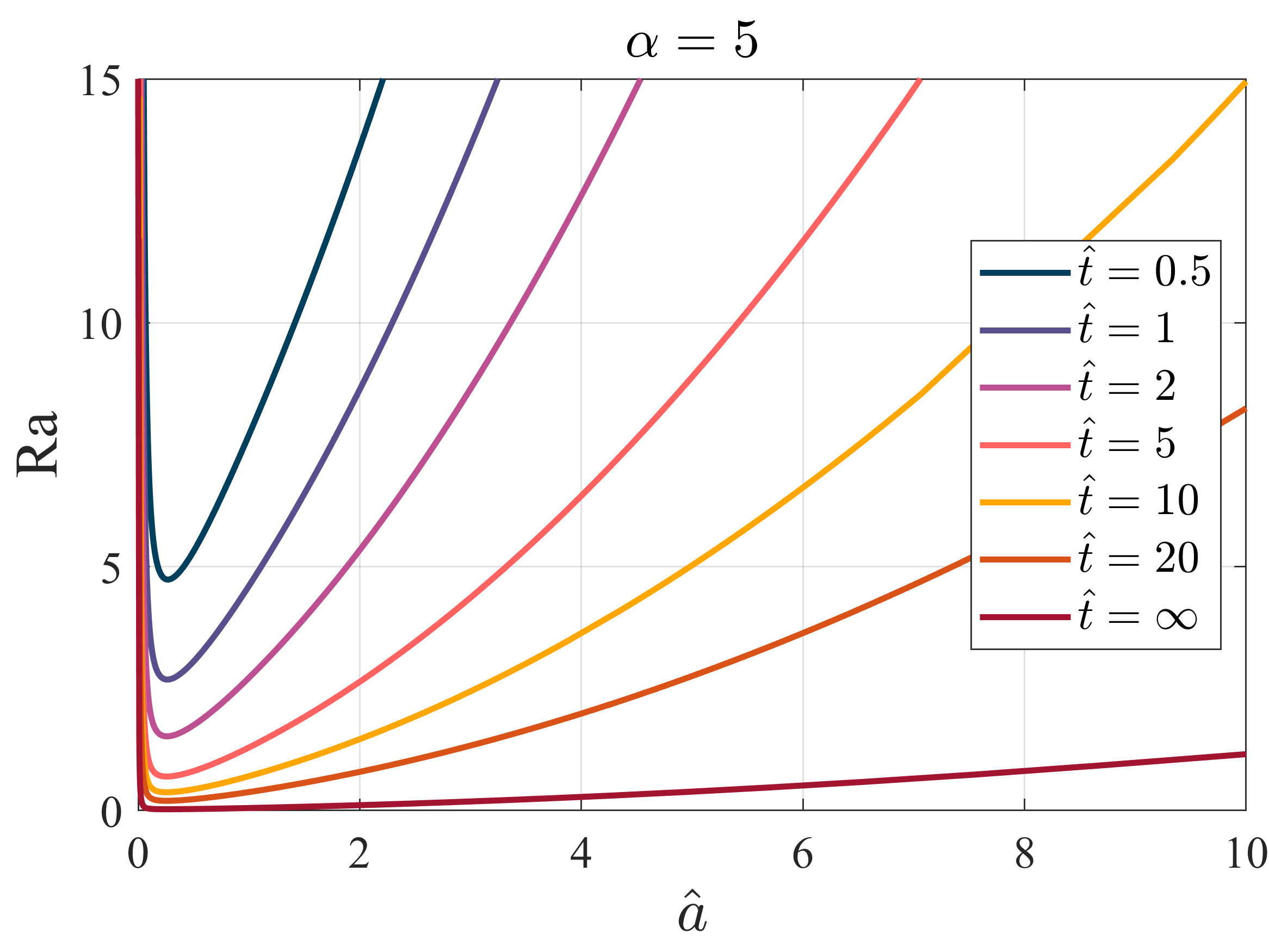}
  \end{minipage}}
\end{figure*}

Solving the eigenvalue problem in section \ref{eigenvalue_problem} yields a Rayleigh number discerning the stable and unstable regime for a given characteristic height, time and perturbation wavenumber. Figure \ref{fig:Ra_over_a} shows this Rayleigh number as a function of $\an$ for $\alpha$ equal to 2 on the left and 5 on the right at several times going from 0.5 to infinity. In order to obtain these results, both solution methods are employed and yield similar results with differences smaller than $10^{-2}$ in the Rayleigh number. 

It is visible that the Rayleigh number corresponding to a fixed wavelength decreases with time, meaning the boundary layer gets more unstable. This is caused by the accumulation of salt at the top of the porous medium over time and the resulting larger differences in liquid density between top and bottom making the system more prone to develop buoyancy-driven instabilities. However, as already alluded to in the ground state section \ref{ground_state_solution}, the fast convergence of the ground-state salinity for small $\alpha$ also leads to a fast convergence of the stability bound against a steady-state bound as visible in the left plot of figure \ref{fig:Ra_over_a} at time $\tn = 10$.

The critical Rayleigh number for the given $\alpha$ and time is equal to the global minimum of the corresponding stability curve. The most unstable perturbation mode has the critical wavenumber $\an_c$ which minimizes $\mathrm{Ra}(\alpha, \tn, \an)$. The first growing perturbation that appears during the temporal evolution of the ground state will have the wavenumber $\an_c$.

\subsubsection{\label{Ra_over_alpha} Critical Rayleigh number as a function of $\alpha$}

\begin{figure}[t!]
\floatbox[{\capbeside\thisfloatsetup{capbesideposition={right,top},capbesidewidth=5.8cm}}]{figure}[\FBwidth]
{\caption[Critical Rayleigh number over $\alpha$]{The figure displays the critical Rayleigh number as a function of $\alpha$ at varying times. The solid lines represent the coinciding eigenvalues that are obtained by the fundamental matrix as well as the Chebyshev-Galerkin method whereas the dashed eigenvalues could only be calculated by the Chebyshev-Galerkin method. The crosses represent the critical Rayleigh numbers calculated by Bringedal et al. \cite{bringedal2022evaporation} for the semi-finite case $\alpha = \infty$.} \label{fig:Ra_over_alpha}}
{\includegraphics[width=6cm]{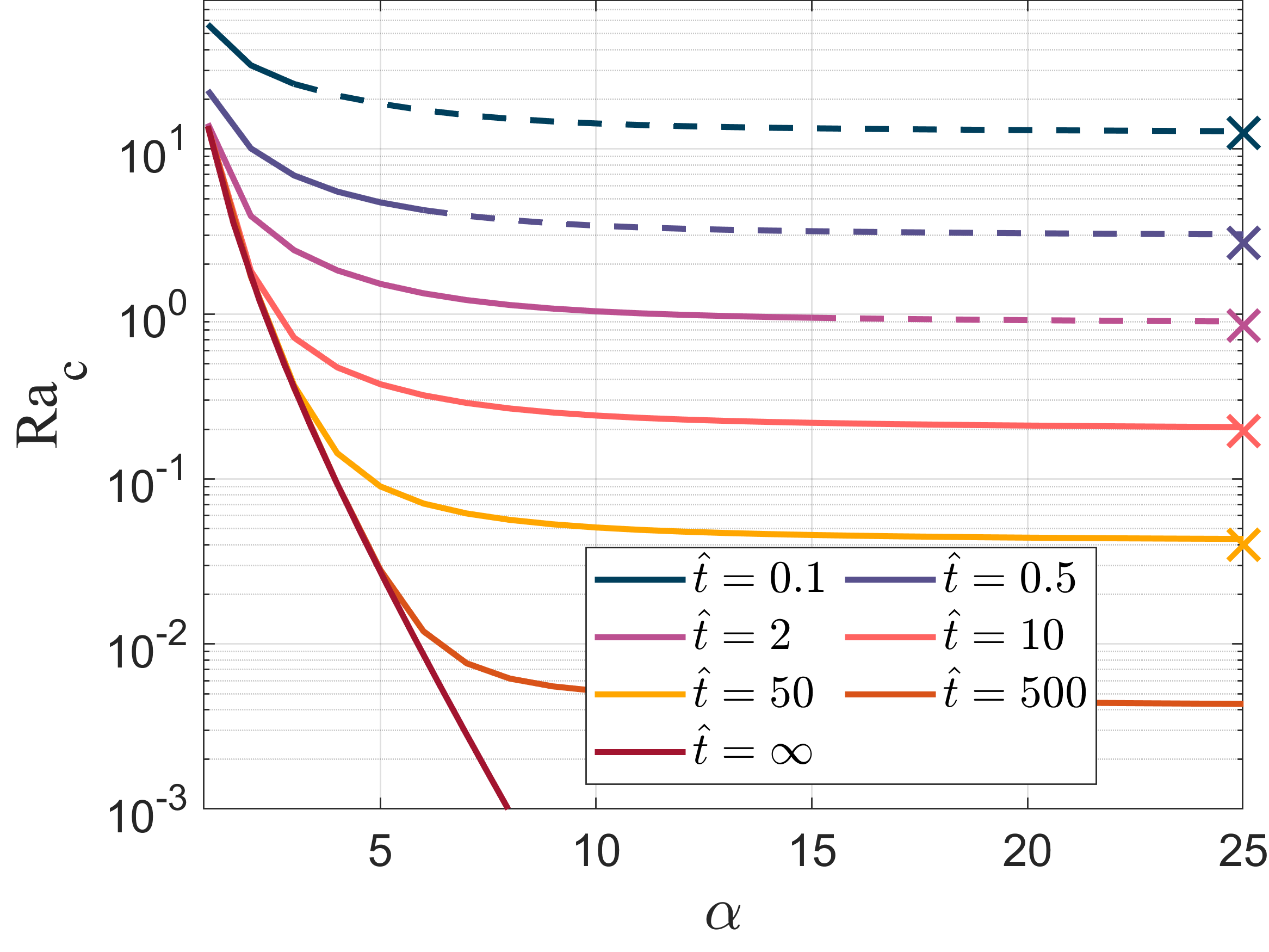}}
\end{figure}

As the critical Rayleigh number $\mathrm{Ra}_c$ is the key quantity determining stability of the system, we will now look at the influence of the dimensionless height on $\mathrm{Ra}_c$. Figure \ref{fig:Ra_over_alpha} shows the critical Rayleigh number as a function of $\alpha$ at different times going from 0.1 to infinity. The solid lines mark the regime where both the fundamental matrix as well as the Cheybshev-Galerkin method produced valid results with differences in the critical Rayleigh number of the order of $10^{-4}$. The dashed lines correspond to critical Rayleigh numbers that could only be calculated with the Cheybshev-Galerkin approach while the fundamental matrix method did not yield eigenvalues. As can be seen in the figure, this only happened for large $\alpha$ and at early times. Under these circumstances, the ground-state salinity exhibits a very sharp increase at the top of the domain, which we have already seen in figure \ref{fig:c_S}, and, thus, its Taylor expansion around the origin has bad convergence behaviour. Since this expansion appears in the series expansion of the system matrix in equation \eqref{system_matrix_coefficients}, the convergence behaviour transfers to the fundamental matrix and the method breaks down. Increasing the cut-off of the series expansion of the fundamental matrix also does not solve the problem as one runs into numerical issues when calculating the coefficient matrices $\mathbf{A_i}$ as defined in equation \eqref{system_matrix_coefficients} for $i$ bigger than 150 due to the factorial term and the i-th derivative $\partial_{\zn}^{(i)} \cn_S$.

The crosses on the right side of figure \ref{fig:Ra_over_alpha} mark the critical Rayleigh number calculated by Bringedal et al. \cite{bringedal2022evaporation} ~at various times. We can see that the critical Rayleigh number for $\alpha$ equal to 25 is already quite close to the semi-infinite case, manifesting quantitatively in a relative difference smaller than 5\%. This shows that both solution methods yield reliable results at large times. For large $\alpha$ and at early times, however, only the Chebyshev-Galerkin approach still provides expected Rayleigh numbers.

Moreover, figure \ref{fig:Ra_over_alpha} shows that for small dimensionless heights, the critical Rayleigh number converges fast against a steady-state value which is in accordance with the fast convergence of the corresponding ground-state salinity. However, for large $\alpha$, for which the underlying ground-state salinity has not converged against its steady state yet, the critical Rayleigh number barely changes with increasing dimensionless height. This is due to the similarity of the ground-state salt concentrations at sufficiently early times which was discussed in section \ref{ground_state_solution} and showed in figure \ref{fig:ground_state_comparison}.

\subsubsection{\label{t_over_Ra} The time of onset}

In figure \ref{fig:Ra_over_alpha}, one can also see that, especially for large $\alpha$, the critical Rayleigh number of the steady state - this is at infinite time - can become very small such that the system is almost guaranteed to become unstable. Consequently, in many physical scenarios, the question is not if but rather when instabilities start to grow. Thus, we will discuss the time of onset of instabilities $\tn_c(\alpha, \mathrm{Ra})$ in this section, which is implicitly defined by
\begin{equation}
    \mathrm{Ra} = \mathrm{Ra}_c(\alpha, \tn_c)~.
\end{equation}
At times $\tn > \tn_c$, the critical Rayleigh number drops below the actual Rayleigh number and, thus, the system becomes unstable.

\begin{figure}[t!]
\floatbox[{\capbeside\thisfloatsetup{capbesideposition={right,top},capbesidewidth=5.8cm}}]{figure}[\FBwidth]
{\caption[Time of onset over Rayleigh number]{The figure shows the time of onset as a function of the Rayleigh number for different $\alpha$. Again, the dashed lines represent the regime where the fundamental matrix method fails to converge and only the Chebyshev-Galerkin method yields reliable results.} \label{fig:t_over_alpha}}
{\includegraphics[width=6cm]{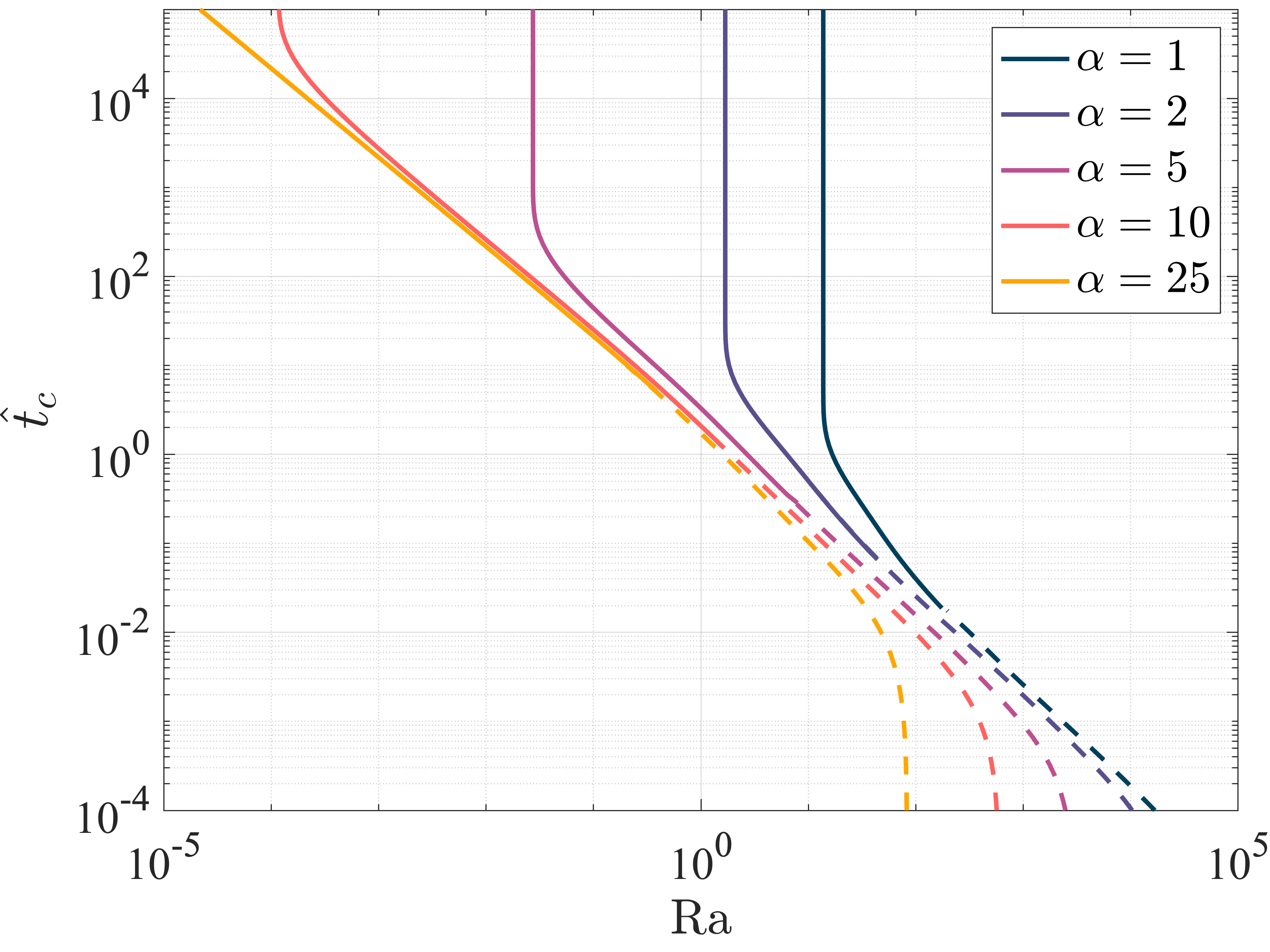}}
\end{figure}

Figure \ref{fig:t_over_alpha} shows the time of onset as a function of $\mathrm{Ra}$ for various $\alpha$ between 1 and 25. Again, the solid lines represent values that could be calculated by both solution approaches whereas the dashed lines are results that could only be produced by the Chebyshev-Galerkin method. Similarly to the previous section, the fundamental matrix method can yield reliable results but fails at small times and for large $\alpha$ due to the bad convergence behaviour of the power series of the ground-state salinity. All the curves for a fixed $\alpha$ in figure \ref{fig:t_over_alpha} have a vertical asymptote at $\mathrm{Ra}_c(\alpha,\tn = \infty)$, which is the critical Rayleigh number once the steady state has been reached. For all $\mathrm{Ra}$ smaller than that, the system will always be stable and the time of onset is infinite. The asymptotes are visible for $\alpha$ from 1 to 10 and one can see that the critical Rayleigh number of the steady state decreases for larger $\alpha$ meaning the system gets more unstable. 

For all $\alpha$, the time of onset decreases monotonically with increasing Rayleigh numbers. Moreover, the time of onset is always decreasing for growing $\alpha$. Hence, enlarging both $\alpha$ and $\mathrm{Ra}$ facilitates the development of instabilities. From this, the definition of $\alpha$ and the Rayleigh number, we can infer the influence of physical parameters on the stability of the boundary layer: A larger diffusion constant and viscosity leads to a more stable setting and delays the onset of instabilities as they either decrease $\alpha$ or $\mathrm{Ra}$ or increase the time scale $\phi D / E^2$. Higher permeability and porosity of the porous medium, a larger height of the porous layer, increased initial salt concentration and initial liquid density all work contrary as they accelerate the development of instabilities and convection.

\section{\label{numerical_solution}Numerical solution}

Up to now we have looked at the system from the point of view of small perturbations to a stable ground-state solution, characterized by a uniform, constant upflow of water. This idea allowed us to clearly distinguish between the stable and unstable part of a solution and calculate the critical Rayleigh number as well as the time of onset of instabilities for a given combination of dimensionless height and Rayleigh number. However, calculating the time-dependent critical Rayleigh number - and thus also the time of onset - via a frozen profile approach of linearized perturbation equations encompasses assumptions which might prove inappropriate. In order to verify that these assumptions are valid, yield reliable results in this scenario and to also investigate the behaviour of perturbations beyond the limits of the linear theory, we run numerical simulation of the original equation system consisting of equations \eqref{nondim_flow_continuity}, \eqref{nondim_darcy}, \eqref{nondim_salt_continuity} together with initial and boundary conditions \eqref{nondim_bc}.

\subsection{\label{numerical_setup}Numerical setup} 

In order to reduce the computational complexity, we set the velocity component $\vn_y$ to zero which allows us to run 2-dimensional instead of 3-dimensional simulations without changing the perturbation behaviour. In this case, the wavenumber $\an$ is now simply only equal to $\an_x$ and the width of the domain is chosen as $2\pi/\an$ with periodic boundary conditions in the lateral direction for all quantities. This way, the time of onset of a single perturbation mode with wavenumber $\an$ can be investigated.

A finite volume scheme is used as discretization method since its conservation property for convection-diffusion problems is useful here. Within the finite volume method, a first-order upwind scheme \cite{moukalled2016finite} is used for the advective flux and a second-order scheme for the diffusive flux \cite{helmig1997multiphase}. The first-order implicit Euler scheme is employed for the time integration of the equations. Since there is a nonlinearity in the advective flux term of the salt conservation equation, we have to solve the arising nonlinear equation system iteratively at every timestep. For that purpose, the salinity in the advection term is initially set to its values at the previous time. This way, all the equations are linear in the unknowns at the new time and a linear equation system can be solved in order to obtain new values. In the next iteration, the newly calculated values of the salinity are used in the advection term and the resulting linear equation system is solved again. This process is repeated until the Euclidean norm of the difference between the salinity, pressure and velocities in the previous iteration and the current iteration is smaller than $10^{-9}$.

In the simulations, 10 finite volume cells with equal width are used in the horizontal direction. This is already sufficient to capture the physical behaviour as only one period of the perturbation has to be resolved. In the vertical direction, however, 360 equisized cells have to be used for $\alpha$ smaller or equal to 5 to produce results which do not change with further refinement of the discretization. The time step size was set to 0.01, such that the grid velocity is higher than the dimensionless evaporation rate $1 / \mathrm{Ra}$.

\subsection{Initial condition}

The value of the salinity is set to zero in all cells in order to adhere to the initial condition as defined in equation \eqref{initial_condition}. However, when investigating the time of onset of instabilities one has to add a small perturbation seed to the initial condition. There are different ways to initialize the perturbations. A simple method is to add a cosine in the lateral direction with wavelength $2\pi / \an$ and amplitude $10^{-10}$ in the top 25 \% of the domain, since the perturbations start growing from the surface. However, such a cosine perturbation needs some time to transform into the shape of a growing perturbation because its amplitude does not have the correct $\zn$-dependence and, thus, tends to overestimate the time of onset.

\begin{figure}[b!]
\floatbox[{\capbeside\thisfloatsetup{capbesideposition={right,top},capbesidewidth=5.8cm}}]{figure}[\FBwidth]
{\caption[Vertical salt-perturbation shape predicted by the linear theory]{The shape of the salt perturbation $\chi$ as predicted by the Chebyshev-Galerkin method is plotted for various parameter combinations. The customized salt perturbation seed in the simulation is then given by $\cn(\xn,\zn) = 10^{-10} ~\chi(\zn) \cos(\an \xn)$.} \label{fig:custom_perturbation}}
{\includegraphics[width=6cm]{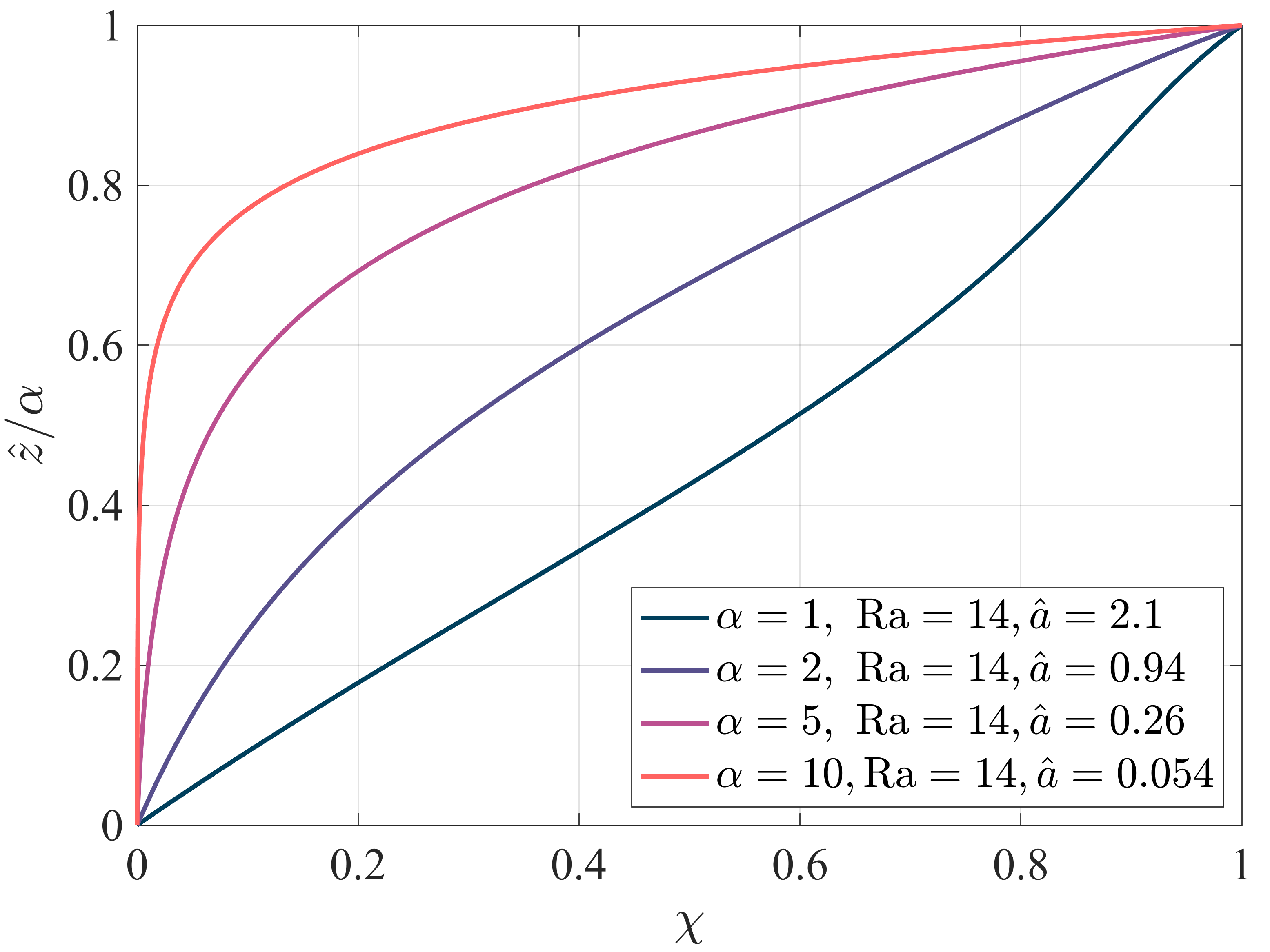}}
\end{figure}

An alternative approach that does not suffer from this problem is initializing the perturbations in the shape that the Chebyshev-Galerkin method predicts. Since the determinant of $\mathbf{M}$ as defined in equation \eqref{matrix_chebyshev} is zero when evaluated at the predicted time of onset, one can calculate the non-trivial kernel of the matrix, which yields a specific shape of the perturbations $\omega$ and $\chi$. Figure \ref{fig:custom_perturbation} shows $\chi(\zn)$ as calculated by the Chebyshev-Galerkin method for several parameter combinations. Initializing the salt concentration with this specifically customized perturbation shape - also with a maximal amplitude of $10^{-10}$ - allows measuring even small times of onset as the perturbation amplitude already has the right $\zn$-dependence. Once initialized, the amplitude of the perturbation is measured as the standard deviation $\sigma_\chi$ of the salt concentration in the first row of cells at the surface. As the ground-state salinity is constant along the $\xn$-direction, this measures the amplitude of the deviation from the ground state at the top of the domain.

\begin{figure}[t!]
  \begin{minipage}[c]{0.49\textwidth}
    \centering
    \includegraphics[width=\textwidth]{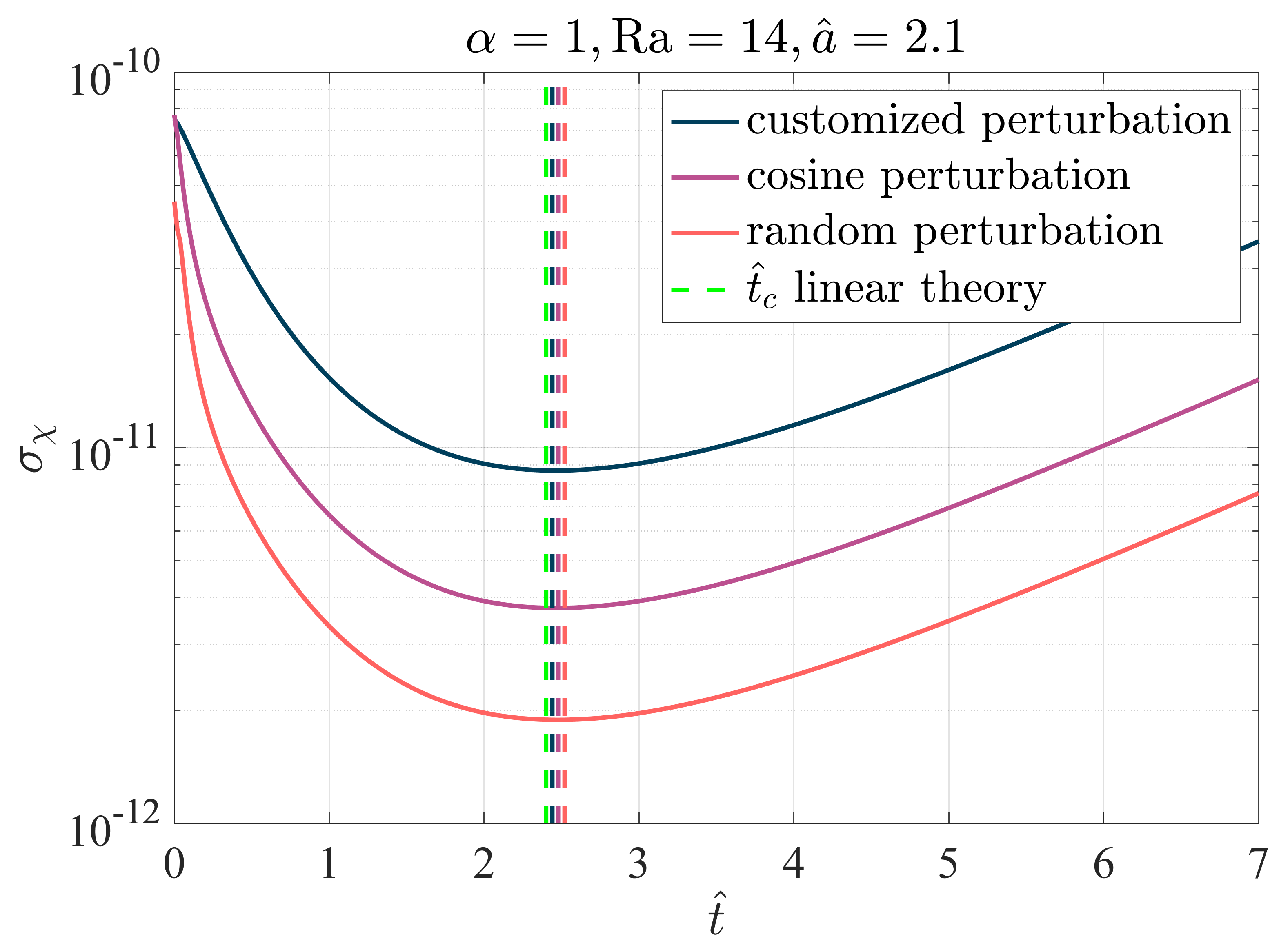}
  \end{minipage}
  \begin{minipage}[c]{0.49\textwidth}
    \centering
    \includegraphics[width=\textwidth]{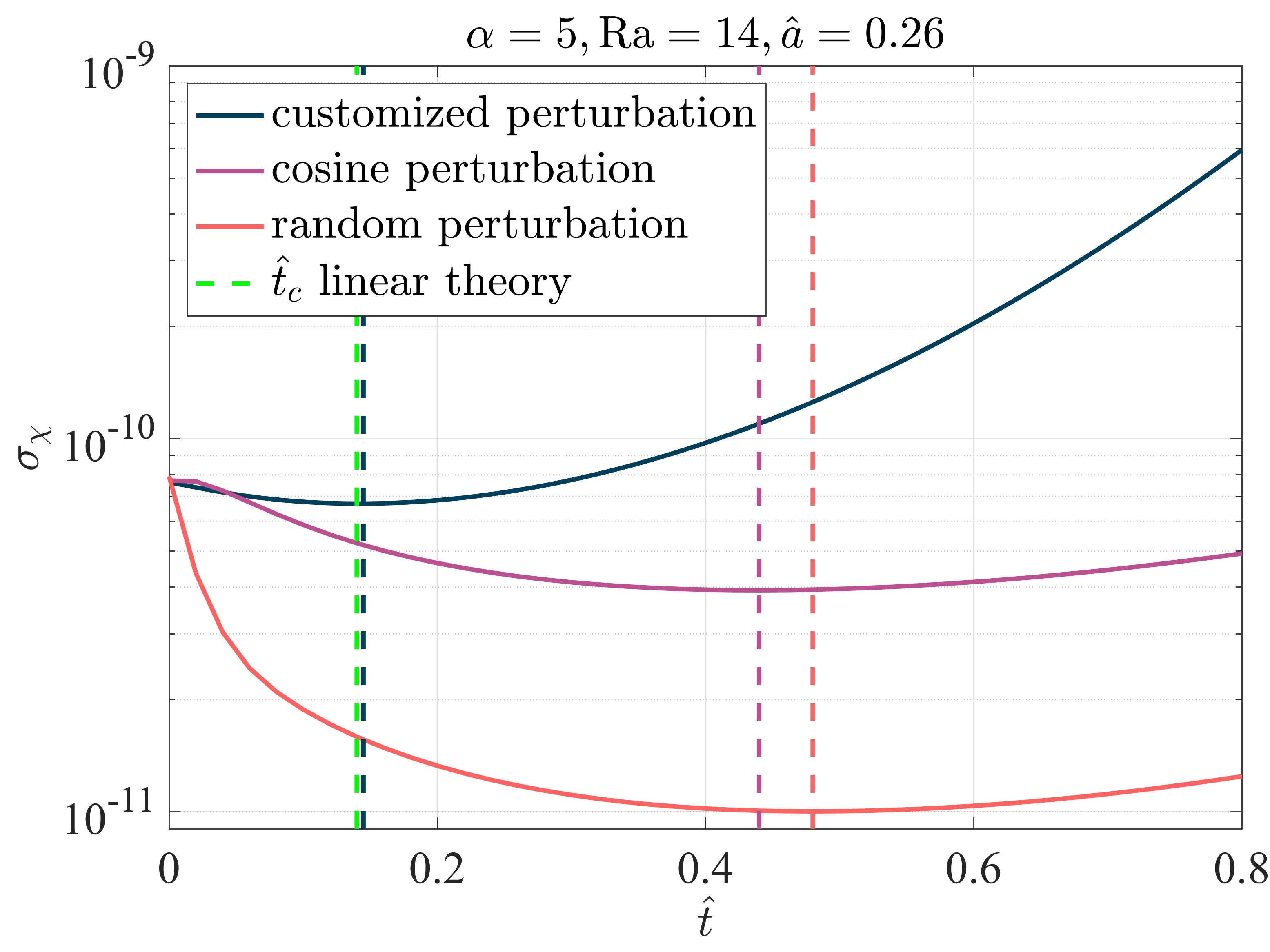}
  \end{minipage}
  \caption[The standard deviation $\sigma_\chi$ for different scenarios]{The standard deviation $\sigma_\chi$ of the salt concentration at the top boundary for different perturbation seeds is plotted over time for two different scenarios. The dotted vertical lines mark the time of onset when the different perturbation seeds are used.}\label{fig:perturbation_comparison}
\end{figure}

Figure \ref{fig:perturbation_comparison} shows an exemplary progress of the standard deviation $\sigma_\chi$ for different perturbation seeds. As a baseline, we also employ a random perturbation which adds a random number drawn from a Gaussian distribution with mean zero and standard deviation $10^{-10}$ to the salinity in the top 25\% of the domain. It is visible that the perturbations are initialized with amplitude of the order of $10^{-10}$ and decay at the beginning until a minimum is reached which corresponds to the onset of convection. In the case of $\alpha = 5$, $\mathrm{Ra} = 14$ and $\an = 0.26$ in the right plot, we can see that the time of onset differs depending on the perturbation seed. Only the specifically customized perturbation coming from the Chebyshev-Galerkin method leads to a time of onset similar as predicted by the linear theory. In the left plot, in contrast, all perturbation seeds yield the same time of onset.

\subsection{\label{ground_state_validation}Validation of the ground state of the simulation}

\begin{figure*}[t!]
  \begin{minipage}[c]{0.45\textwidth}
    \centering
    \includegraphics[width=\textwidth]{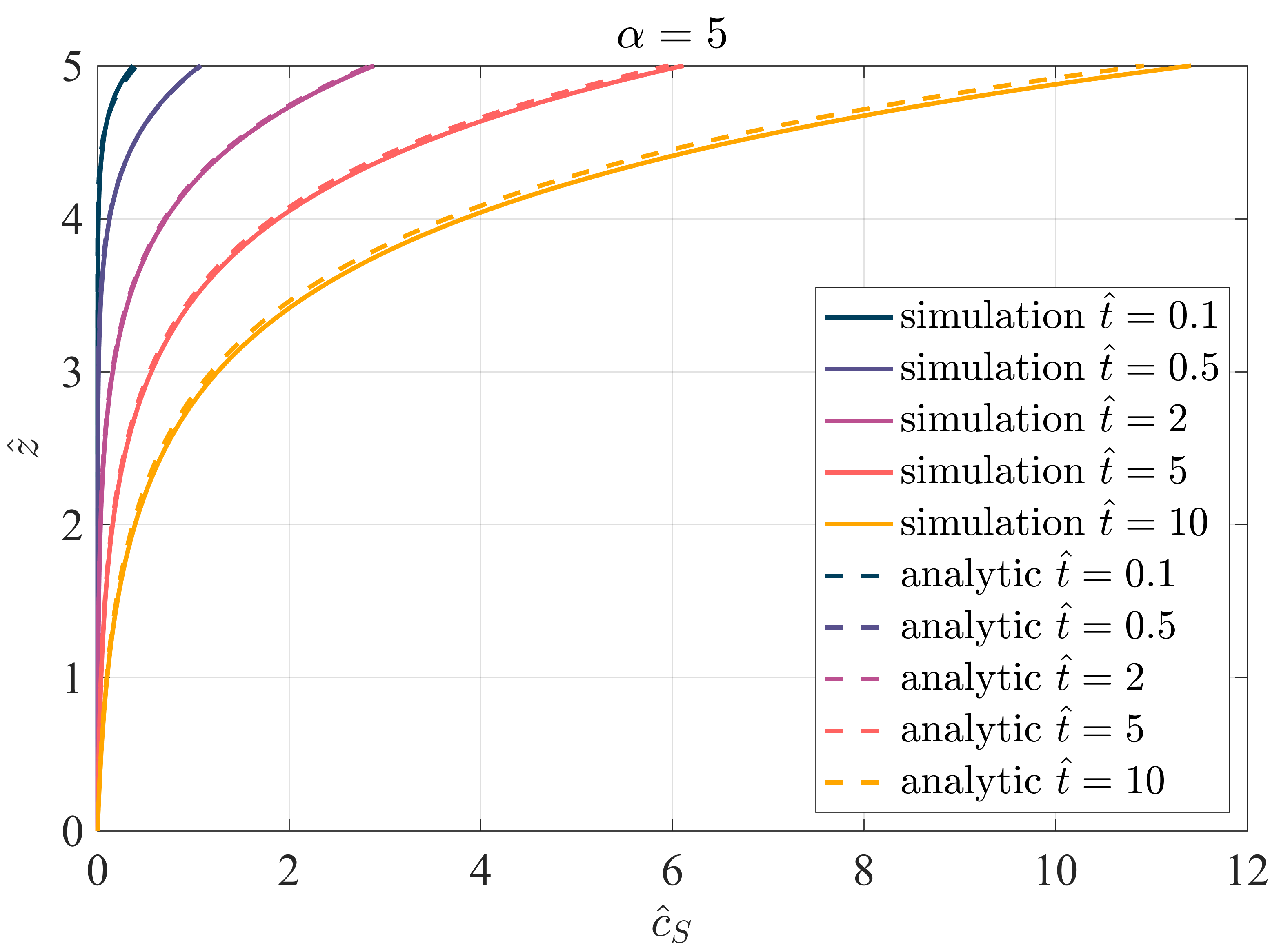}
  \end{minipage}
  \begin{minipage}[c]{0.45\textwidth}
    \centering
    \includegraphics[width=\textwidth]{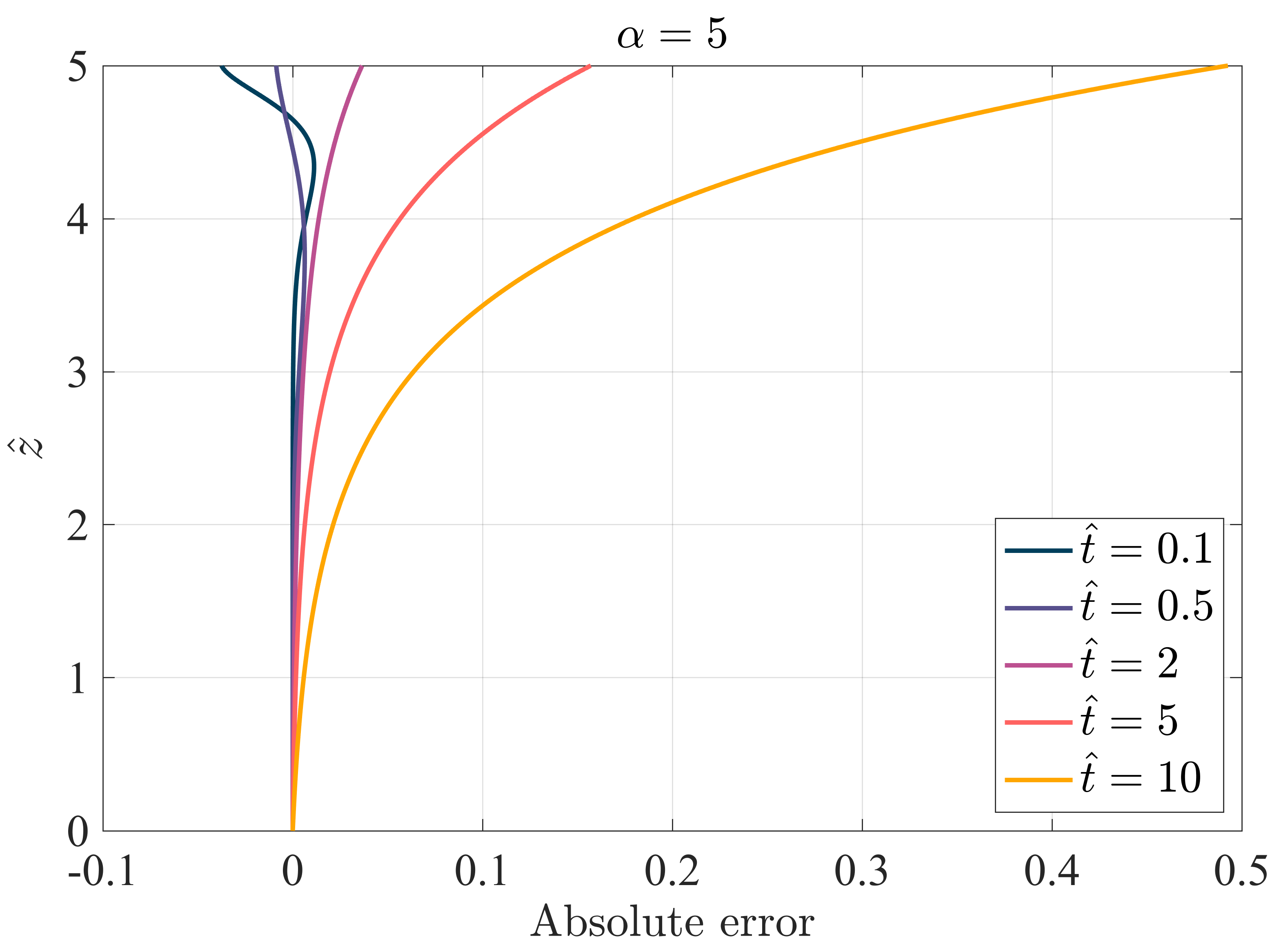}
  \end{minipage}
  \caption[Comparison of the numerically calculated and analytic ground-state salt concentration]{The left plot depicts the analytic as well as the numerically obtained ground-state salt concentration, which was calculated by the 2-dimensional simulation, for $\alpha = 5$. The right plot shows the absolute difference between both solutions. }\label{fig:dns_analytic}
\end{figure*}

The numerical setup can be validated by comparing the vertical ground-state salt concentration profile to the analytic solution, which was derived in section \ref{ground_state_solution}. By averaging the salt concentration calculated by the simulation over the $\xn$-direction, the vertical ground-state profile is obtained. The left plot of figure \ref{fig:dns_analytic} shows the ground-state salinity of the simulation in comparison to the analytic solution from equation \eqref{c_S} for $\alpha$ equal to 5. The right plot of figure \ref{fig:dns_analytic} displays the absolute difference between simulation and analytic solution. It can be seen that the simulation is able to reproduce the behaviour of the analytic ground state to a high degree of accuracy. Even at the top of the domain, where the absolute error is largest at all times, the relative error of the simulation ground state is still below 5\%.

\subsection{\label{comparison_time_of_onset}Comparison of the time of onset to linear theory}

The time of onset in the direct simulation is measured for $\alpha$ equal to 1, 2 and 5 at a Rayleigh number of 3 and 14 as these representative values arise from reasonable values of the underlying physical quantities. The wavenumbers of the perturbation modes are set to the critical wavenumber predicted by the linear theory for each dimensionless height, being $\an = 2.1$ for $\alpha = 1$, $\an = 0.94$ for $\alpha = 2$ and $\an = 0.26$ for $\alpha = 5$.

\begin{table}[b!]
    \centering
	\caption[Time of onset: linear theory vs. numerical simulation]{The times of onset as predicted by the linear theory and as measured in the simulation using different perturbation seeds are listed for several scenarios. The dashes in the $\alpha = 1$ and $\mathrm{Ra} = 3$ case mean that the system stays stable at all times.}
	\label{tab:time_of_onset}
	\vspace{0.5cm}
	\begin{adjustbox}{center}
	\begin{tabular}{c|cc|cc|cc}
		& \multicolumn{2}{c|}{$\alpha = 1, \an = 2.1$} & \multicolumn{2}{c|}{$\alpha = 2, \an = 0.94$} & \multicolumn{2}{c}{$\alpha = 5, \an = 0.26$} \\[2pt]
		& $\mathrm{Ra} = 3$ \quad \quad& $\mathrm{Ra} = 14$ & $\mathrm{Ra} = 3$ \quad \quad& $\mathrm{Ra} = 14$ & $\mathrm{Ra} = 3$ \quad \quad& $\mathrm{Ra} = 14$ \\\hline
		linear theory &-\quad \quad& 2.44 &3.05\quad \quad& 0.31 & 0.87 \quad \quad& 0.14 \\
	    custom perturbation &-\quad \quad& 2.42 &3.04\quad \quad& 0.33 & 0.84\quad \quad & 0.14 \\
		cosine perturbation &-\quad \quad& 2.42 &3.04\quad \quad& 0.46 & 1.54\quad \quad& 0.44 \\
		random perturbation &-\quad \quad& 2.42 & 3.08 \quad \quad& 0.47& 1.49 \quad \quad & 0.29 \\
	    \bottomrule
	\end{tabular}
	\end{adjustbox}
\end{table}

Table \ref{tab:time_of_onset} contains the time of onset of the linear theory as well as the time of onset measured in the numerical simulations using the customized perturbation, the cosine perturbation and the Gaussian random perturbation as seed. The simulation results when using the customized perturbation seed corroborate the time of onset of the linear theory for all considered cases as the largest relative difference is below 7\%. The cosine and random perturbation seed both lead to an overestimation of the time of onset for several cases. Their time of onset is only in accordance with the linear theory when the initial perturbation has enough time before the onset is expected to turn into an exponentially growing shape. For the cases $\alpha = 1, \mathrm{Ra} = 14$ and $\alpha = 2, \mathrm{Ra} = 3$ the cosine perturbation yields the same, and the random perturbation almost the same result as the customized perturbation seed. In both these scenarios, there is enough time before the time of onset to turn the initial perturbation into the right shape. When the onset time is early, as in the case for $\alpha=2,\mathrm{Ra}=14$ and $\alpha=5$, the deviations between the onset times are larger, and only the custom perturbation is close to the onset time predicted by the linear theory.

\subsection{\label{full_convection_patterns}Nonlinear convection patterns}

\begin{figure}[b!]
  \begin{minipage}[c]{0.48\textwidth}
    \centering
    \includegraphics[width=\textwidth]{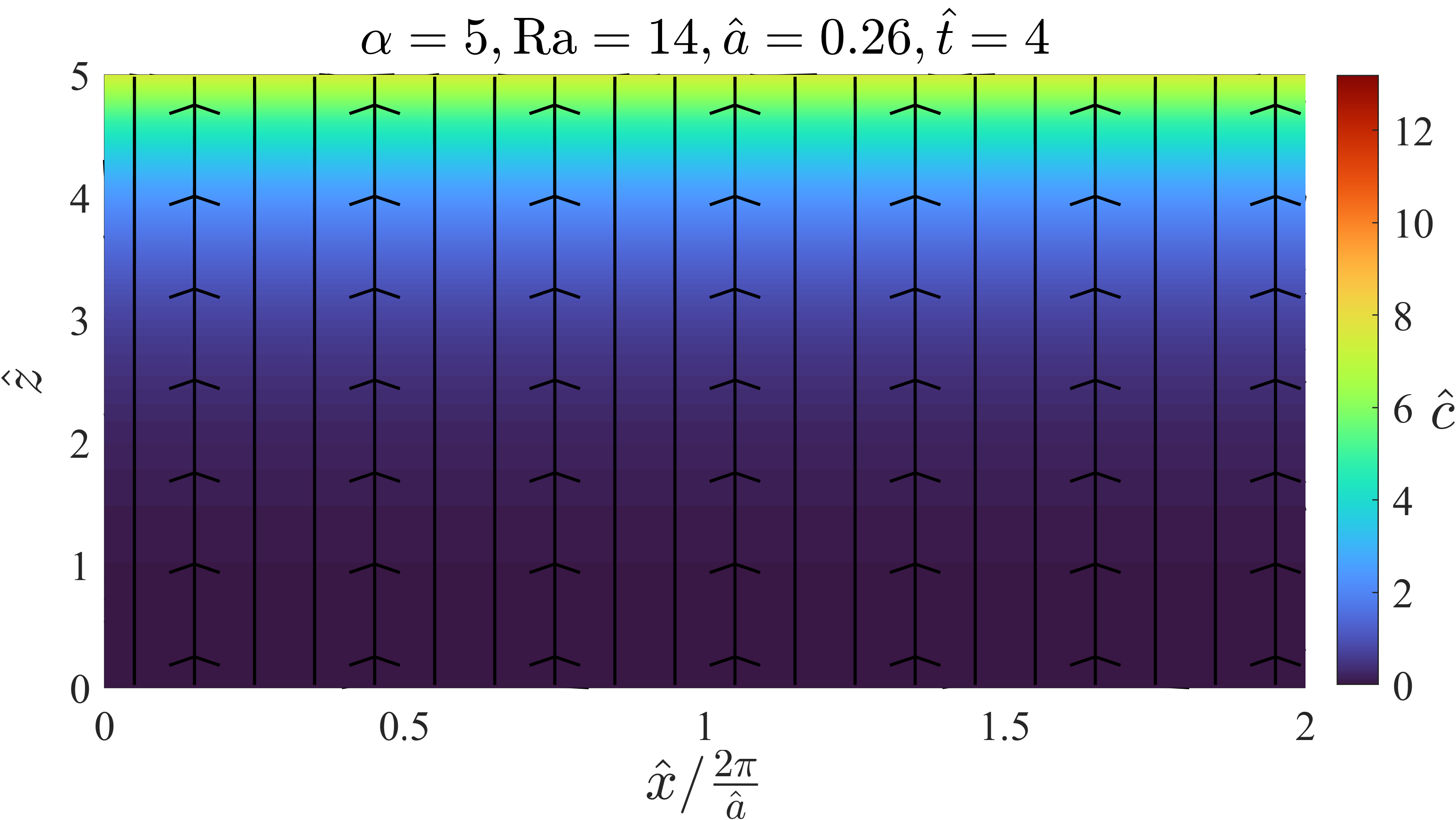}
  \end{minipage}
  \begin{minipage}[c]{0.48\textwidth}
    \centering
    \includegraphics[width=\textwidth]{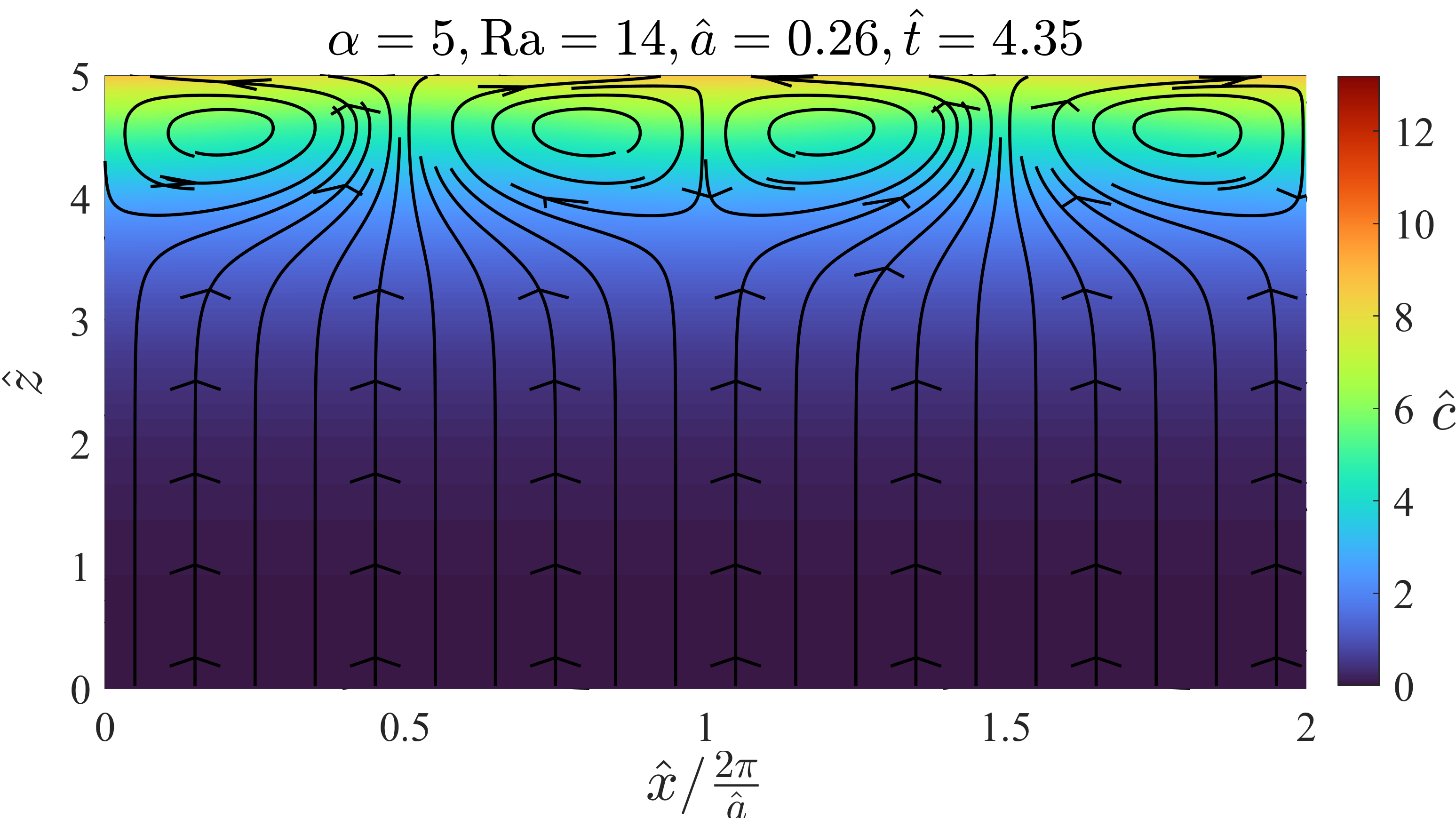}
  \end{minipage}
  \caption[Simulation result at time A and B]{The left plot visualizes the system for $\alpha = 5$, $\mathrm{Ra} = 14$ and $\an = 0.26$ at $\tn = 4$ (time A), which is right before the instabilities become noticeable. Here, however, the deviation from the ground state is barely visible as the streamlines are still all vertical. The right plot shows the same system at $\tn = 4.35$ (time B). At the top, vortices are appearing as there are regions of high salinity where fluid is flowing downwards. Times A and B are marked in figure \ref{fig:cmax}.}\label{fig:convection1}
\end{figure}

After the onset of convection, the instabilities are growing exponentially in strength until the perturbations have a sufficiently large amplitude such that nonlinear effects are not negligible anymore. 
The left plot in figure \ref{fig:convection1} shows the simulation result for $\alpha = 5$, $\mathrm{Ra} = 14$ and $\an = 0.26$ at $\tn = 4$ (time A in figure \ref{fig:cmax}), which is right before the nonlinear effects start to dominate the behaviour. At this time, we can see that the flow is still mainly characterized by a homogeneous upwards flow and the salt concentration does not visibly change in the lateral direction. Hence, the deviations from the ground state are small enough such that they are not apparent yet. 

\begin{figure}[t!]
  \begin{minipage}[c]{0.48\textwidth}
    \centering
    \includegraphics[width=\textwidth]{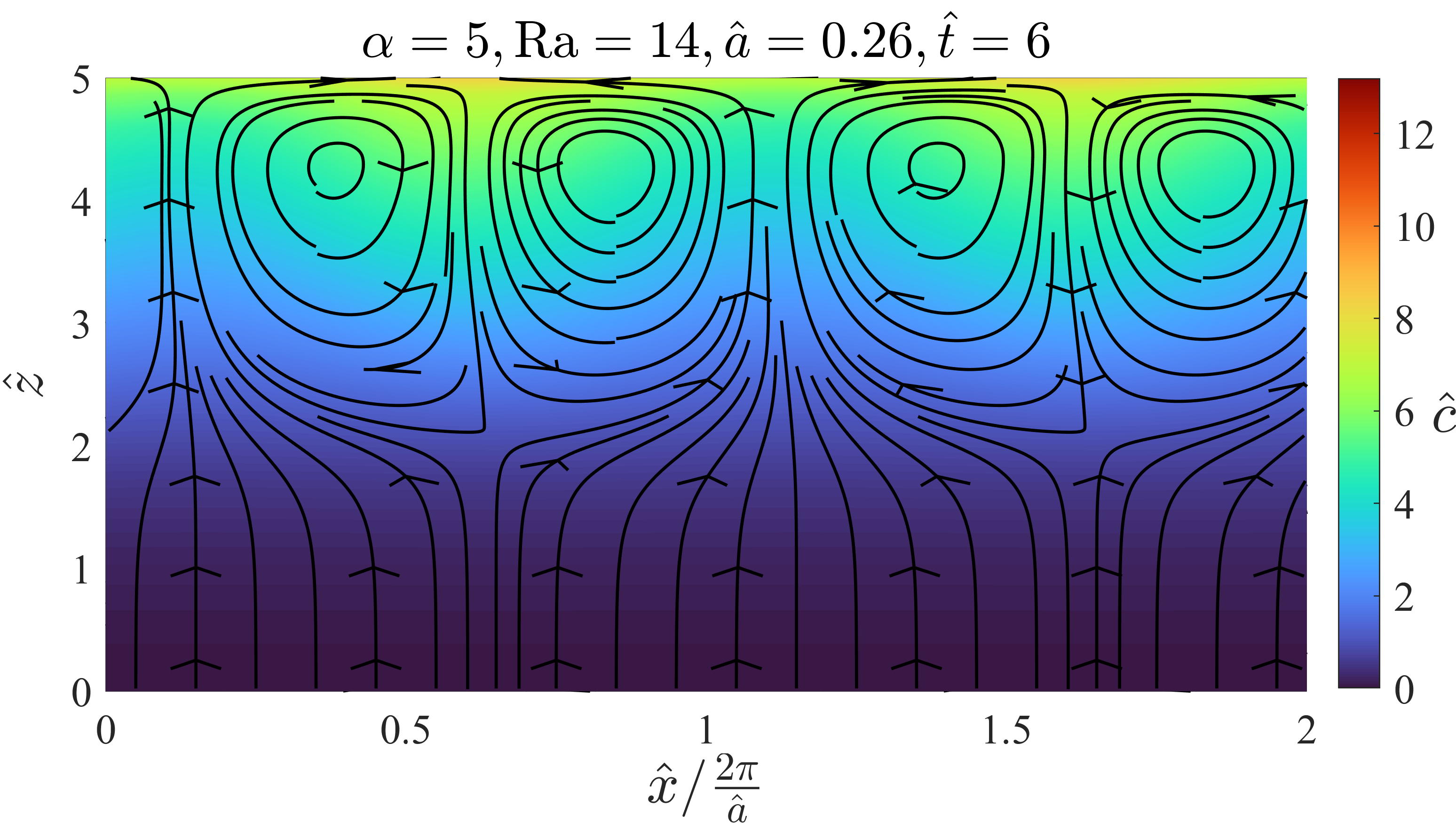}
  \end{minipage}
  \begin{minipage}[c]{0.48\textwidth}
    \centering
    \includegraphics[width=\textwidth]{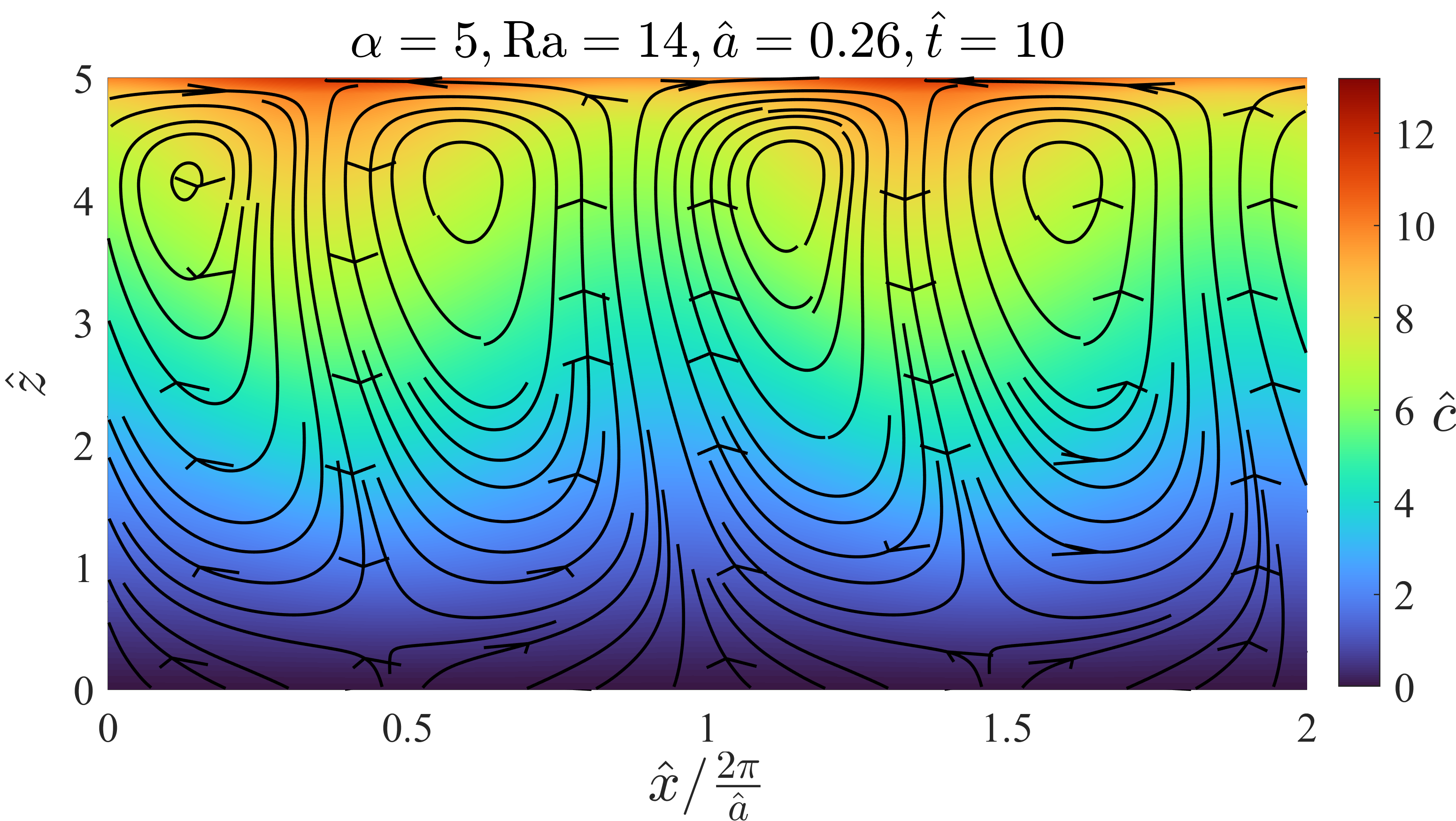}
  \end{minipage}
  \caption[Simulation result at time C and D]{The left plot shows the same system as in figure \ref{fig:convection1} at $\tn = 6$ (time C), where the vortices already cover half of the domain and visibly transport salt towards the bottom of the porous slab. The right plot depicts the system at $\tn = 10$ (time D). Here, the convection vortices have reached the bottom of the porous slab and salt fingers have started to become clearly visible. Times C and D are marked in figure \ref{fig:cmax}. }\label{fig:convection2}
\end{figure}

The right plot in figure \ref{fig:convection1} shows the same system slightly later at $\tn = 4.35$ (time B in figure \ref{fig:cmax}). Here, the variations in the salt concentration at the top of the porous slab have led to the development of convection vortices. As there are alternating regions of higher and lower salt concentration at the top of the slab due to the instabilities, there are differences in buoyancy. Hence, the liquid starts to flow downwards in the areas of high salinity and continues flowing upwards in between, resulting in the convection vortices. As the vortices distribute salt from the surface to lower parts of the domain, the maximum salinity decreases for a short period, which can be seen in figure \ref{fig:cmax}. At this time, however, these vortices are still confined to a small portion of the domain as the flow in the lower half of the slab still looks like the ground-state flow.

\begin{figure}[b!]
\floatbox[{\capbeside\thisfloatsetup{capbesideposition={right,top},capbesidewidth=5.8cm}}]{figure}[\FBwidth]
{\caption[Simulation result at time E]{This plot shows the same system as figures \ref{fig:convection1} and \ref{fig:convection2} at $\tn = 20$ (time E). Now, it is visible that the salt concentration near the bottom has got significantly larger due to the salt transport by the salt fingers. Time E is marked in figure \ref{fig:cmax}.} \label{fig:convection3}}
{\includegraphics[width=6cm]{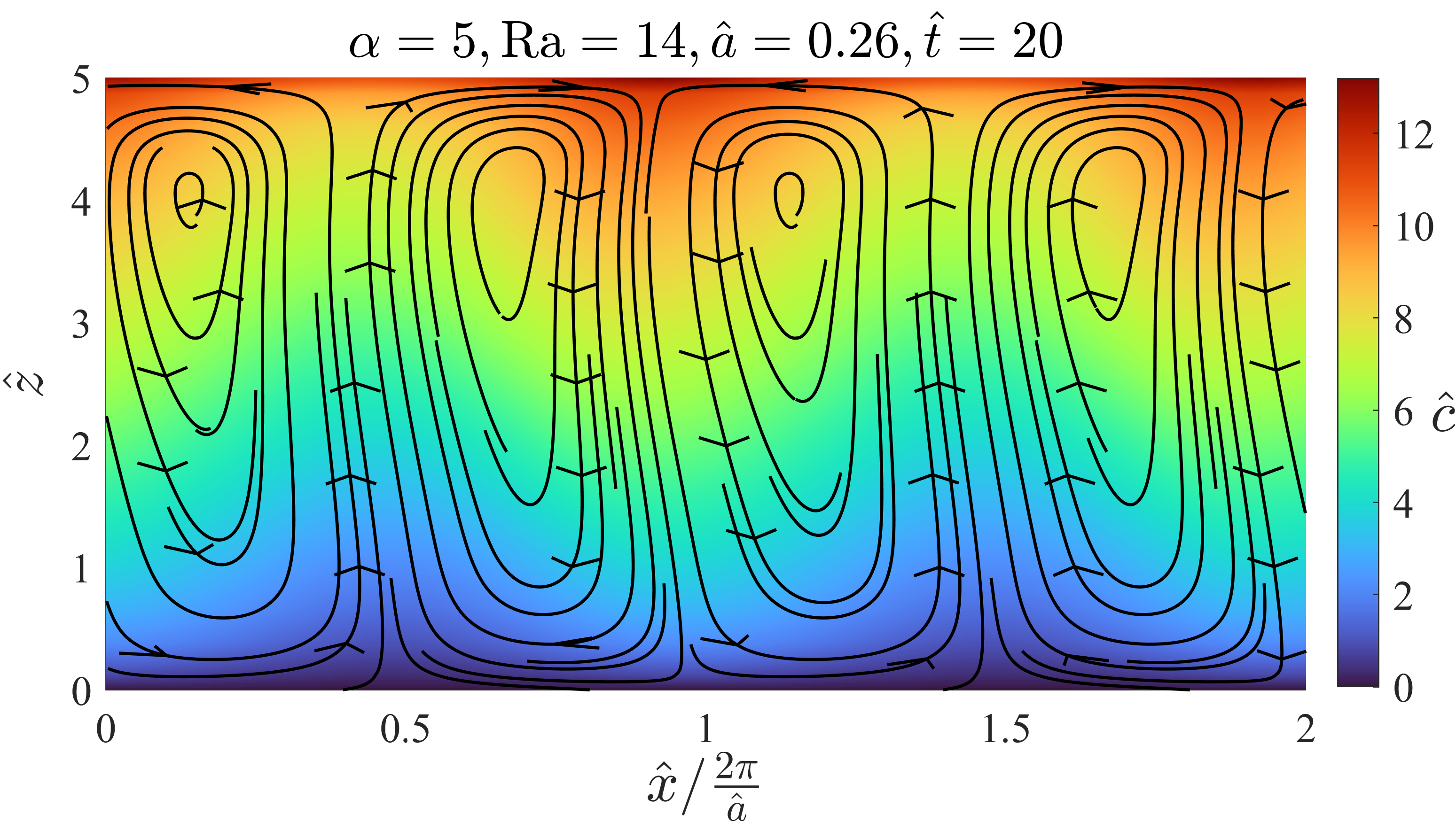}}
\end{figure}

The left plot in figure \ref{fig:convection2} depicts the system at $\tn = 6$ (time C in figure \ref{fig:cmax}), where the vortices have grown from the top and now expand to the middle of the domain. The region, in which convective flow occurs, is visibly more saline than the lower part of the porous medium. This is due to the downwards transport of salt from the top by the convection vortices. As the vortices do not span the entire domain yet, the salt transported downwards by them does not leave the porous slab and is instead partially transported upwards again. Hence, the salt concentration at the top is again increasing at this time as figure \ref{fig:cmax} shows.

The right plot in figure \ref{fig:convection2} shows the system at $\tn = 10$ (time D in figure \ref{fig:cmax}), where the vortices begin to span the entire vertical extent of the domain. Due to the transport of salt along the regions of downwards flow, the salt fingers are starting to become more clearly visible when looking at the salt concentration profile. As the Dirichlet boundary conditions for salt and flow velocity at the bottom boundary prevent the salt fingers from growing any further here, the salt starts to accumulate at the lower part of the salt fingers. The increasing salt concentration gradient near the bottom boundary leads to more diffusive flux of salt out of the domain.

Figure \ref{fig:convection3} displays the system at $\tn = 20$ (time E in figure \ref{fig:cmax}). As the salt concentration gradient near the bottom boundary has increased even further, enough salt diffuses out of the domain to counteract the advective salt inflow prescribed by the bottom boundary conditions. Hence, there is zero net flux of salt into the domain and the salt concentration at the top does not increase any further.
This is also visible when looking at figure \ref{fig:cmax}, which shows the evolution of the maximum salinity $\cn_\mathrm{max}$ in the entire domain as measured in the simulation over time. Before $\tn = 4$, $\cn_\mathrm{max}$ is almost the same as in the ground state, since the perturbations are still small in amplitude. When the perturbations become large enough, convection vortices occur and consequently the maximum salinity falls rapidly as salt is transported away from the top and more evenly distributed through the domain. However, as the vortices only span a portion of the domain, salt does not leave the domain and is instead transported to the top again. Thus, the maximum salinity is growing again after the convection vortices have built, which is also in accordance with observations by Bringedal et al. \cite{bringedal2022evaporation} for the vertically unbounded case. Only when the convection vortices have reached the bottom of the porous medium at $\tn = 10$, the diffusive flux of salt out of the domain is gradually increasing. Finally, in the yellow marked regime in figure \ref{fig:cmax}, there is an equilibrium of diffusive outflow and advective inflow of salt at the bottom boundary, such that the maximum salinity does not grow anymore.

\begin{figure}[t!]
\floatbox[{\capbeside\thisfloatsetup{capbesideposition={right,top},capbesidewidth=5.8cm}}]{figure}[\FBwidth]
{\caption[Development of the maximum salt concentration over time in a convective scenario]{This plot depicts the temporal development of the maximum salt concentration $\hat{c}_\mathrm{max}$ as measured in the simulation (black line) in comparison to the ground state (gray line). Moreover, the times A-E at which the state was visualized in figures \ref{fig:convection1}, \ref{fig:convection2} and \ref{fig:convection3} are marked.} \label{fig:cmax}}
{\includegraphics[width=6cm]{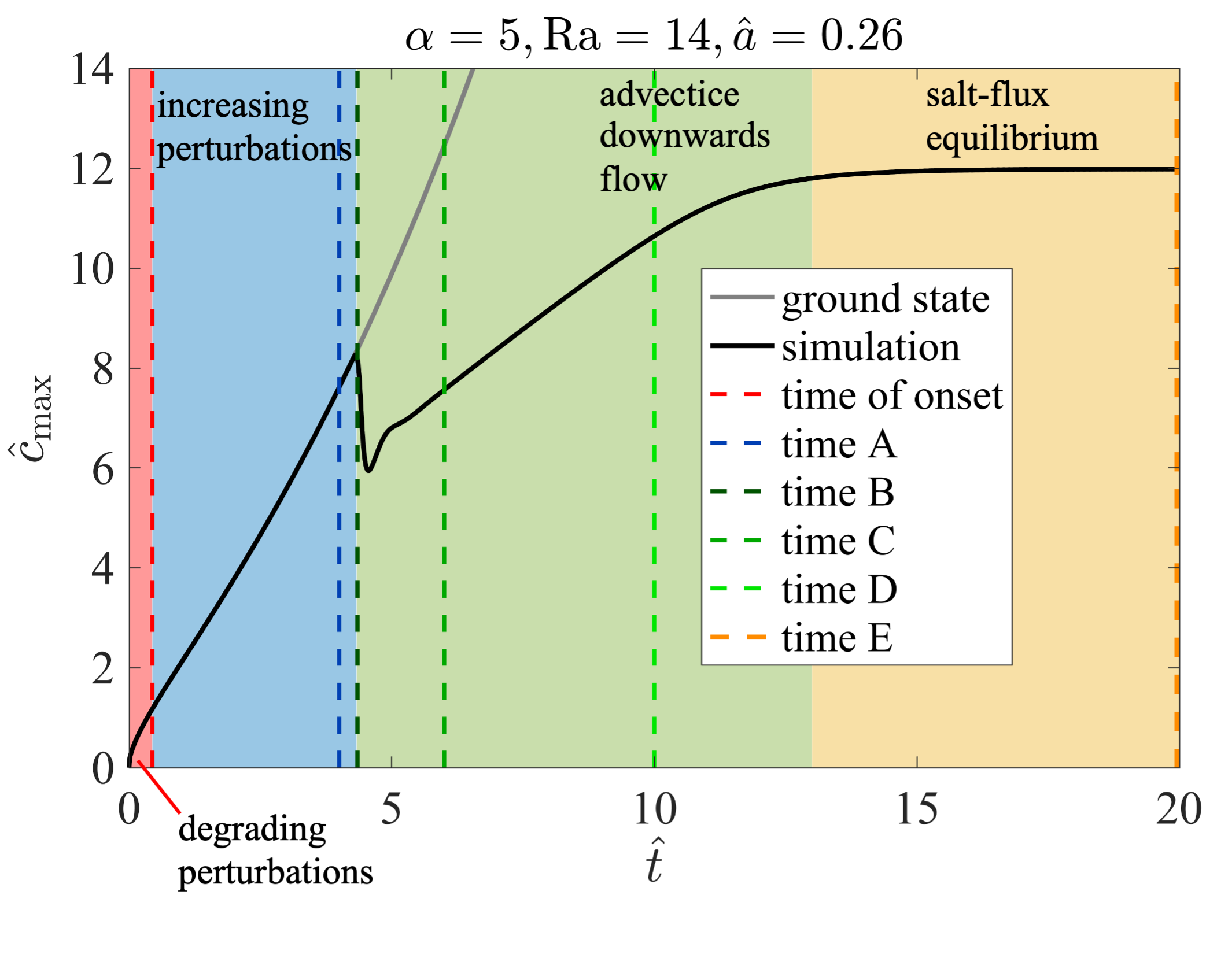}}
\end{figure}

Even though we have only considered a specific scenario in this section, the general behavior of the development of the instabilities is similar for other dimensionless heights, Rayleigh numbers and perturbation wavenumbers. Slim \cite{slim2014solutal} has decomposed the dynamics of solutal convection in a porous medium into several regimes, which can also be done in our scenario: First, perturbations are degrading until the time of onset is reached (red regime in figure \ref{fig:cmax}). This corresponds to the "diffusive regime" of Slim\cite{slim2014solutal}. Then, the perturbations are growing exponentially according to linear stability theory (blue regime in figure \ref{fig:cmax}), which Slim calls the "linear growth regime"\cite{slim2014solutal}. When instabilities, and thus the variations in the salt concentration at the surface are large enough, regions of advective downwards flow develop at the top of the porous slab and expand downwards. This regime of advective downwards flow in figure \ref{fig:cmax} corresponds to the "flux-growth regime" of Slim\cite{slim2014solutal}. Now, the salt carried downwards by the vortices starts to accumulate near the bottom of the salt fingers. The large salinity gradient due to the fixed salt concentration at the bottom of the porous medium leads to a increasing diffusive flux out of the domain. Once a salt-flux equilibrium between advective inflow and diffusive outflow is reached at the bottom boundary, the maximum salinity stops growing. This is analogous to the "shut-down regime" of Slim\cite{slim2014solutal} as the salt concentration profile also reaches a steady state.

\section{\label{conclusions}Conclusions}

In this study, the onset of convection in a laterally unbounded liquid-saturated slab of porous medium with finite height was investigated. The isotropic and homogeneous porous medium is coupled to a reservoir of saline water at the bottom. The water is flowing upwards due to an evaporation-induced throughflow. A no-flux boundary condition is used for the salt at the top of the porous medium leading to the gradual accumulation of salt at the surface. 

As long as no buoyancy-driven instabilities occur, the evolution of the dimensionless ground-state salt concentration is entirely determined by the dimensionless height of the porous slab. In the stable regime, this height corresponds to the P\'eclet number and only depends on the evaporation rate $E$, the height of the slab $H$ and the effective diffusion constant $D$. 

The ground-state salt concentration was derived analytically within the framework of Sturm-Liouville theory. The stability of this ground state was investigated by a linear stability analysis. Herein, a Chebyshev-Galerkin method as well as a novel fundamental matrix method were employed in order to solve the perturbation eigenvalue problem. The fundamental matrix approach depends on a power series expansion of the system matrix describing the perturbation system. Thus, it runs into problems for ground states with a slowly converging power series as the numerical evaluation of the power series becomes inaccurate. When the power series of the ground state exhibits sufficiently fast convergence behaviour, however, the fundamental matrix method is highly accurate and was shown to be in accordance with the Chebyshev-Galerkin method. The fundamental matrix method itself can be applied to a wide range of stability problems in different areas of physics. 

At large dimensionless heights, the critical Rayleigh numbers obtained by the linear stability analysis were shown to be in good agreement with the results by Bringedal et al. \cite{bringedal2022evaporation}, who considered the case of a semi-infinite porous slab. Moreover, the time of onset of convection was calculated as a function of the dimensionless height and the Rayleigh number. Here, the time of onset was shown to decrease with increasing Rayleigh number and dimensionless height, thus, making the system more prone to convection. The times of onset measured in the direct numerical simulation corroborate the results obtained by the linear stability analysis, when using a numerical perturbation mimicking the perturbation from the linear stability analysis. 

By running simulations for longer times, the development of nonlinear convective flow and its influence on the salt concentration could also be investigated. The evolution of the system can be decomposed into different regimes analogous to the work of Slim\cite{slim2014solutal} on solutal convection in porous media. Before the time of onset, the system is in the regime of degrading perturbations. Afterwards, the instabilities grow according to the linear stability theory in a regime of increasing perturbations. Once their amplitude is large enough, nonlinear effects lead to convection vortices and salt fingers growing from the top in a regime of advective downwards flow. Finally, when the salt fingers expand to the bottom of the porous slab, the regime of salt-flux equilibrium is reached, where there is zero net flux of salt into the porous slab. Hence, the salt concentration comes to a steady state in which the salinity at the surface is smaller than in the stable, non-convective case. That means, when triggered early enough, the instabilities can prevent salt precipitation and formation of salt crust formation at the top of the porous medium.


\section*{Acknowledgments}
Funded by the Deutsche Forschungsgemeinschaft (DFG, German Research Foundation) - Project Number 327154368 - SFB 1313.

\section*{Author declarations}
\noindent\textbf{Conflict of interest}\\ The authors declare that they have no confict of interest.

\noindent\textbf{Data availability}\\ The codes used to solve the eigenvalue problem and the governing system of equations are available in  https://doi.org/10.18419/darus-3057.

\noindent\textbf{Author contributions}\\ \textbf{L. Kloker}: Software, Investigation, Methodology (lead), Validation, Visualization, Writing/Original Draft Preparation, Writing/Review \& Editing (lead). \textbf{C. Bringedal}: Conceptualization, Funding Acquisition, Methodology (supporting), Supervision, Writing/Review \& Editing (supporting).

\bibliographystyle{plain} 
\bibliography{paper}

\end{document}